\documentclass[fleqn,usenatbib]{mnras}

\usepackage{newtxtext,newtxmath}

\usepackage[T1]{fontenc}

\DeclareRobustCommand{\VAN}[3]{#2}
\let\VANthebibliography\thebibliography
\def\thebibliography{\DeclareRobustCommand{\VAN}[3]{##3}\VANthebibliography}

\usepackage{graphicx}	
\usepackage{amsmath}	
\usepackage{diagbox}
\usepackage{csquotes}
\usepackage{ dsfont }
\usepackage{bm}
\usepackage{xcolor}
\usepackage{comment}
\usepackage{upgreek}

\title[Numerical convergence of MRI simulations]{On the numerical convergence of MRI simulations}

\author[T. Jannaud and H. N. Latter]{
Thomas Jannaud\thanks{E-mail: taj30@cam.ac.uk}
and Henrik N. Latter
\\
DAMTP, University of Cambridge, CMS, Wilberforce Road, Cambridge CB3 0WA, UK
}

\date{Accepted 2025 November 6. Received 2025 October 31; in original form 2025 April 11}

\defcitealias{Fromang2007}{FP07}

\pubyear{\the\year{}}

\begin{document}
\label{firstpage}
\pagerange{\pageref{firstpage}--\pageref{lastpage}}
\maketitle

\begin{abstract}

The magnetorotational instability (MRI) plays a crucial role in the evolution of many types of accretion disks. It is often studied using ideal-MHD  numerical simulations. In principle, such simulations should be numerically converged, i.e. their properties should not change with resolution. Convergence is often assessed via the MRI quality factor, $Q$, the ratio of the Alfvén length to the grid-cell size. If it is above a certain threshold, the simulation is deemed numerically converged. In this paper we argue that the quality factor is not a good indicator of numerical convergence. First, we test the performance of the quality factor on simulations known to be unconverged, i.e. local ideal-MHD simulations with zero net-flux, and show that their $Q$s are well over the typical convergence threshold. The quality-factor test thus fails in these cases. Second, we take issue with the linear theory underpinning the use of $Q$, which posits a constant vertical field.
This is a poor approximation in real nonlinear simulations, where the vertical field can vary rapidly in space and generically exhibits zeros. We calculate the linear MRI modes in such cases and show that the MRI can reach near-maximal growth rates at arbitrarily small scales. Yet, the quality factor assumes a single and well-defined scale, near the Alfvén length, below which the MRI cannot grow. We discuss other criticisms and suggest a modified quality factor that addresses some, though not all, of these issues.


\end{abstract}




\section{Introduction}

Accretion disks are found around many types of astrophysical objects: black holes (AGNs, \cite{Padovani2017} or black hole binaries, \cite{Tetarenko2016}), stars \citep{benisty2023}, neutron stars \citep{MunosDarias2014}, white dwarfs \citep{Zorotovic2020}, and even planets \citep{Benisty2021}. The accretion onto the central object is powered by both external torques (most notably involving magnetized jets and winds) and internal turbulent torques \citep{Ferreira1995,Lesur2021Winds,Pascucci2023,Lesur2023PP7}. The preferred driver of turbulence in most accretion disks is the magnetorotational instability \citep[MRI;][]{Balbus1991,Balbus1998}, as it only requires two ingredients: a weak magnetic field and a rotation speed decreasing with radius. The MRI has been studied extensively through magnetohydrodynamic (MHD) numerical simulations \citep[see][for a review]{Lesur2021}.

MRI simulations can be organized by three characteristics: (a) geometry - local, cylindrical, or global; (b) ideal or non-ideal MHD; and (c) net or zero-net magnetic flux . Local simulations use the shearing box approximation. Suitable for small-scale phenomenona in thin disks, it conveniently expresses the MHD equations in cartesian coordinates \citep{Goldreich1965,Latter2017}, but introduces artificial symmetries. Ideal simulations neglect viscosity and resistivity, and while some accretion disk processes can be described in quasi-ideal MHD, turbulent cascades do need (small-scale) dissipation; additionally, sufficiently strong non-ideal effects not only dampen the MRI turbulence but also signficantly modify its large-scale behaviour \citep[e.g.,][]{kunz2013,lesur2014, Bai2014,Bethune2016}. The net-flux configuration usually assumes the presence of a mean vertical magnetic field over the whole numerical domain. While it allows for the launching of magnetized jets and winds in global and vertically stratified local models \citep[e.g.][]{suzuki2014,fromang2013,Lesur2013,bs2013}, and as we will see eases convergence of the simulations, the origin and properties of the assumed large-scale magnetic structure are not always constrained. On the other hand, \cite{Jacquemin2024} have shown the emergence of large-scale vertical field from zero-net-flux initial conditions, but it is weak compared to the turbulent small-scale field.

For the results extracted from such simulations to be deemed reliable, one needs to know if they are numerically converged, i.e. if their properties are independent of the grid size. The simplest and cleanest way to assess numerical convergence is to run the simulation at several resolutions. As there is no resistivity or viscosity in ideal simulations, a better resolved simulation will always extend the turbulent cascade to smaller scales. Nonetheless, one can check that large-scale or mean quantities (e.g. the normalized stress, $\alpha$) do not vary significantly with the grid-cell size. This method has been used to prove the non-convergence of local zero-net-flux (ZNF) ideal MHD simulations (for certain aspect ratios): \cite{Pessah2007} and \cite{Fromang2007} (hereafter FP07) showed that $\alpha \propto 1/N$, where $N$ is the number of grid cells along one direction. This was confirmed by several later numerical studies \citep{Simon2009,Guan2009,Bodo2011}. It was long thought that the issue of non-convergence in ZNF boxes could be solved by vertical stratification \citep{Davis2010,Hawley2011}. However, \cite{Bodo2014} and \cite{Ryan2017} numerically uncovered a weaker dependence on resolution ($\alpha \propto N^{-1/3}$), and so they are unconverged as well.

Convergence tests are numerically costly, as they require additional simulations at better resolution. The aforementioned simulations are relatively simple local models, and for more demanding global setups it is hard to fit both the science runs and the convergence tests within one's allocated computing time. This motivated the search for a convergence metric that would obviate the need to run more demanding simulations. The most popular of those metrics is the quality factor $Q_z$ \citep{Noble2010,Hawley2011}, defined as the ratio of the local vertical Alfvén length to the vertical grid-cell size, averaged over the simulation domain. Using ZNF stratified simulations of the MRI, \citet{Hawley2011} prescribed certain thresholds above which $Q_z$ must lie for a simulation to be judged numericaly converged, and then applied these tests to global simulations \citep[see also][]{Hawley2013}. A convergence study in `semi-global' cylindrical models was performed by \cite{Sorathia2012}, who derived analogous convergence criteria involving both the vertical quality factor $Q_z$ and the toroidal quality factor $Q_\upvarphi$. 

Subsequently, these criteria have become a common diagnostic of numerical convergence in both global  MHD simulations \citep{Hawley2013,suzuki2014,Zhu2018,Gagnier2024} and global GRMHD simulations \citep{McKinney2014,White2019,Kiuchi2022,Hayashi2023,Scepi2024}, to cite only a few papers. However, given that the ZNF stratified runs underpinning \citet{Hawley2011} have since been shown to be unconverged, the quality factor deserves a re-appraisal. 
In this paper we argue that the MRI quality factor is, in fact, a flawed metric of convergence. It is useful to establish if an initial condition yields sufficiently fast-growing MRI modes \citep[e.g.][]{sano2004,flock2010} but, in the ensuing turbulent nonlinear state, it at best offers necessary but insufficient conditions for convergence. 

We present a set of general arguments criticising the theoretical underpinnings of the MRI quality factor (section \ref{sec:TheoreticalBackground}), and then give two detailed illustrative calculations. First, in section \ref{sec:Simulation}, we test the quality factor criteria on simulations that we know are unconverged, ZNF shearing boxes. If the quality factor is a reliable diagnostic, we expect $Q_z$ to return low values in these simulations, but instead we find uncomfortably large values. Next, in section \ref{sec:LinearTheory}, we interrogate the linear theory underlying the quality-factor criteria, which rely on a fixed and well-defined physical scale for the MRI. Motivated by the prevalence of current sheets in MRI-induced turbulence \citep{sano2007,zhdankin17,ross2018,kawazura24}, and thus regions where the vertical field may change sign, we undertake a simple linear MRI calculation in a radially varying vertical field.
Our analysis shows that, if the background field changes sign, normal modes can grow on arbitrarily small scales at near the maximum rate: thus, there is no reliable and fixed physical scale that a quality factor can be anchored upon. Finally, in section \ref{sec:Conclusions}, we discuss a revised definition of $Q_z$, which uses the (usually weak) background vertical field threading disk annuli, and how it might be usefully employed in global simulations.

\section{Theoretical background}\label{sec:TheoreticalBackground}

In this section, we start by reiterating some of the necessary theoretical background of the linear MRI, provide the basic definition of the quality factor, and summarise its usage in recent numerical simulations. We then briefly discuss general objections to the quality factor, when used as a diagnostic of numerical convergence, before our more detailed calculations in later sections. 


\subsection{Linear MRI: Vertical mean-field configuration}\label{sec:LinearMRI}

We consider a local patch of a Keplerian disk in the incompressible shearing box approximation. 
The disk is initially in its laminar background state, rotating at a rate $\Omega$, with a uniform density $\rho$ and magnetic field $\bm{B}_0 = B_0 \bm{e}_z$. For sufficiently small axisymmetric vertical harmonic perturbations of the form $e^{i k_z z + s t}$, we get the usual local dispersion relation of the MRI:
\begin{equation}\label{eq:DispersionRelationMeanField}
    D(s,V_{A} k_z) = s^4 + s^2\left( \Omega^2 +2 V_{A}^2 k_z^2 \right) + V_{A}^2 k_z^2 \left( V_{A}^2 k_z^2 - 3 \Omega^2 \right) = 0,
\end{equation}
where $V_{A}^2 = B_0^2 / (4 \pi \rho)$ is the squared Alfvén speed \citep[see derivation in][for example]{Balbus1991}. Remarkably, equation (\ref{eq:DispersionRelationMeanField}) is also a good approximation in compressible and vertically stratified boxes \citep{Gammie1994, Latter2010}. It describes a linear instability when $s^2$ is real, which is when $k_z < \sqrt{3} \Omega /V_{A}$. This means that for a given magnetic field $B_0$ there is a critical wavenumber above which there are no MRI modes. One can frame this in terms of wavelength:
\begin{equation}\label{eq:MeanFieldCriticalLambda}
    \lambda > \lambda_\text{crit} (B_0) = \frac{2 \pi \lvert V_{A} \rvert}{\sqrt{3} \Omega} = \frac{\sqrt{\pi} \lvert B_0 \rvert}{\sqrt{3 \rho} \Omega},
\end{equation}
where $\lambda = 2 \pi / k_z$. All modes of wavelength below $\lambda_\text{crit}$ cannot grow.

\subsection{The quality factor: Definition and motivations}\label{sec:DefinitionQualityFactor}

The existence of such a critical wavelength is incredibly useful to numericists. Given a constant background field $B_0$, if the  grid-cell size is sufficiently small with respect to $\lambda_\text{crit}(B_0)$, then we can be sure all linear axisymmetric MRI modes are resolved \citep[e.g.][]{Hawley2011}.

This motivated the (vertical) quality factor, which was to our knowledge first introduced by \cite{Noble2010} and developed into the currently used tool by \cite{Hawley2011}. It is defined as
\begin{equation}\label{eq:DefinitionQz}
    Q_z = \sqrt{3} \frac{\lambda_\text{crit}(B_z)}{\delta z} = \frac{2 \pi \lvert V_{Az} \rvert}{\Omega \delta z} = \frac{2 \pi l_{Az}}{\delta z} = \frac{\sqrt{\pi} \lvert B_z \rvert}{\sqrt{\rho} \Omega \delta z},
\end{equation}
where $\delta z$ is the grid-cell size in the vertical direction and $l_{Az} = \lvert V_{Az} \rvert / \Omega$ is the vertical Alfvén length, with $V_{Az} = B_z / \sqrt{4 \pi \rho}$ the local vertical Alfvén speed, calculated at any time during the simulation. Thus, crucially, and in contrast to section \ref{sec:LinearMRI}, $V_{Az}$ depends on the local vertical magnetic field $B_z$, and not the global background magnetic field $B_0$; it thus comprises both the background field and any turbulent field that arises in the simulation. It is both a function of space and time, though often some average is taken. Note that the factor $\sqrt{3}$ is to make the numerator close to the wavelength of the fastest growing mode, $(3\sqrt{5}/4) \lambda_\text{crit}$.

Obviously, if the simulation starts with a uniform vertical field throughout the domain, then for the early evolution of the disk, the dispersion relation (equation \ref{eq:DispersionRelationMeanField}) might hold approximately in each local region of the domain. The $Q_z$ would then provide a measure for how well the linear MRI modes are resolved in the growing phase. 

The leap is then to assume that, once the MRI saturates, the characteristic scale of the ensuing turbulence, e.g.\ its injection scale, is $\sim \lambda_\text{crit}(B_z)$, 
with $B_z$ the field generated by the self-same turbulence, not the initial field (as mentioned above).
This is a kind of `quasi-linear' theory of MRI saturation, which has no firm theoretical basis as of yet \citep[though see][]{ebrahimi2009}. But if it is granted, then $Q_z$ provides a measure for how well the maximum (linear) turbulent injection scale is resolved.

Finally, the vertical quality factor is often used alongside a toroidal one, also introduced by \cite{Hawley2011} and defined as $Q_y = 2 \pi l_{Ay} / \delta y$, for shearing boxes, and $Q_\upvarphi = 2 \pi l_{A\upvarphi} / (R\delta \upvarphi)$, in global domains (with $R$ radius). Here $l_{Ay}$ and $l_{A\upvarphi}$ are the local and global toroidal Alfven lengths derived from the relevant toroidal field in each case, $B_y$ or $B_\upvarphi$, and $\delta y$ is the grid-cell size in the azimuthal direction of the shearing box, and $\delta \upvarphi$ is the grid-cell size in azimuth in global models.


\subsection{The use of quality factors in the literature}\label{sec:HowItsUsedInLit}

The vertical quality factor $Q_z$ is the most commonly used, as it is derived directly from the resolvability of the linear MRI. If it is over a certain threshold, the simulations are deemed converged. Using results from ZNF vertically stratified simulations, \citet{Hawley2011} recommended a threshold between 10 and 20:
\enquote{\itshape Although $\alpha$ increases dramatically when $Q_z$ rises past a few, its dependence on resolution (in relative terms) appears to level out
in the range $10 \lesssim Q_z \lesssim 20 $.}
 In addition \cite{Hawley2011} suggests taking $10$ as a threshold for the toroidal quality factor. 
 
 A limited number of global convergence studies followed. Based on their cylindrical simulations, \cite{Sorathia2012} stated a requirement of $Q_z \geq 10-15$ for small toroidal quality factors ($Q_\upvarphi \simeq 10$) and a requirement of only $Q_z \geq 6$ for when $Q_\upvarphi \simeq 25$. \cite{Hawley2013}, on the other hand, recommended thresholds of $Q_z \geq 15$ and $Q_\upvarphi \geq 20$, from fully global set-ups.


These metrics are commonly used to assess the numerical convergence of simulations, local or global (cylindrical or spherical), relativistic or not, stratified or unstratified, net-flux or zero-net-flux. Both vertical and toroidal quality factors are most often averaged over the whole domain \citep{Bai2013,McKinney2014,Zhu2018}. However, in the presence of vertical stratification, $\lvert V_{A_z} \rvert$ quickly rises with altitude; far away from the midplane, the MRI is suppressed but the quality factor sharply increases \citep[see last paragraph of section 3.1. in][]{Hawley2011}. As a consequence, some works compute the midplane value of $Q_z$ \citep{suzuki2014,Takaso2018}.

In addition, \cite{Sorathia2012} introduced the convergence indicators
\begin{equation}\label{eq:Resolvability}
    \mathcal{R}_{z} = \frac{1}{V} \int_V \mathcal{H}(Q_z-8) dV, \hspace{0.5cm} \mathcal{R}_{\upvarphi} = \frac{1}{V} \int_V \mathcal{H}(Q_\upvarphi-8) dV,
\end{equation}
where $\mathcal{H}$ is the Heaviside step function. $\mathcal{R}_z$ is the vertical `resolvability', defined as the fraction of the simulation volume $V$ where $Q_z \geq 8$, and $\mathcal{R}_\upvarphi$ is the toroidal `resolvability', defined as the fraction of $V$ where $Q_\upvarphi \geq 8$. If these factors are sufficiently high, the MRI is thought to be resolved in a large enough fraction of the simulation, and the simulations are judged globally converged. Empirically, simulations with $\mathcal{R}_z \geq 0.7$ and $\mathcal{R}_\upvarphi \geq 0.8$ are often deemed converged \citep{Sorathia2012,Parkin2013,Gagnier2024}. 

In summary, studies employing the quality factor define a `converged' simulation state as one that exhibits quality factors over a given threshold. They then check that the averaged quality factors in their simulations are over this threshold ($\left< Q_z \right>_V \geq 15$ and $\left< Q_\upvarphi \right>_V \geq 20$); or that their minima are above it ($\text{min}_V\left( Q_z \right) \geq 15$ and $\text{min}_V\left( Q_\upvarphi \right) \geq 20$); or that the cells with quality factors under the threshold are in the minority ($\mathcal{R}_z \geq 0.7$ and $\mathcal{R}_\upvarphi \geq 0.8$).

\subsection{Issues with the quality factor}\label{sec:IssuesQualityFactor}

Several objections could be leveled at the quality factor, when used to assess the numerical convergence of MRI simulations. Before animating some of these criticisms with example calculations in sections \ref{sec:Simulation} and \ref{sec:LinearTheory}, we  briefly discuss them below.

\subsubsection{The MRI's local and global linear theory and $Q_y$/$Q_\upvarphi$}\label{sec:Qphi}

To begin, subthermal background azimuthal fields play a negligible role in the \emph{local axisymmetric} MRI \citep{blaesbalb94}, and thus $Q_y$ is unimportant for its resolvability. However, in the absence of any vertical field, the azimuthal field is important for weakly and transiently growing \emph{non-axisymmetric} waves (the so called `toroidal MRI'). But when the vertical field is restored these shearing waves are completely transformed, and far more vigorous \citep{Balbus1992,squire2014}. Thus, unless $B_z=0$ (and thus $Q_z=0$!), it cannot be claimed that $Q_y$ measures the resolvability of the local non-axisymmetric MRI either. 

In global or quasi-global disks, the non-axisymmetric MRI manifests very differently, and now possesses well-defined growth rates \citep{Curry1996}\footnote{Note that strong azimuthal fields quench the axisymmetric global (or quasi-global) MRI, but permit additional axisymmetric instabilities \citep{curry95,Pessah2005,Das2018}}. The properties of these modes are more complicated than their axisymmetric counterparts. For instance, in the presence of a purely toroidal background field, \citet{Terquem1996} and \citet{Ogilvie1996} showed that the fastest growing modes appeared in the limit $k_z \rightarrow \infty$, or $\lambda \rightarrow 0$.  \cite{Brughmans2024} show that, in a mixed toroidal and vertical background field, modes with negative toroidal wavenumbers generally possess a smaller critical vertical wavelength than axisymmetric modes (equation \ref{eq:MeanFieldCriticalLambda}), as previously noted by \cite{Begelman2022}. Moreover, there appears to be no radial scale below which some non-axisymmetric modes cease to grow, or at least that scale is not close to $l_{A \upvarphi}$. Thus, it seems $Q_\upvarphi$ cannot decide on the resolvability of global linear non-axisymmetric modes, in general.

In fact, the arbitrarily small scales upon which the non-axisymmetric MRI grows might appear to invalidate $Q_z$ as a good metric for the resolvability of the linear MRI in total, in all its guises. It must be said that, though many papers discuss $Q_\upvarphi$ in the context of resolving the `linear toroidal MRI', \citet{Hawley2011} ultimately conclude that $Q_\upvarphi$ might best be thought of measuring non-axisymmetric \emph{nonlinear} processes within MRI turbulence and the MRI dynamo.


\subsubsection{The linear MRI in an unsteady and inhomogeneous magnetic field}\label{sec:SpaceVaryingMagneticFields}

Let us consider a MRI simulation in its saturated phase. The local magnetic field $B_z$ strongly varies in space (see Fig. \ref{fig:CoupeBy} or \ref{fig:NFCoupeBz}). But when deploying the quality factor, $|B_z|$ is often averaged in all or part of the simulation domain. This averaged field is then promoted to the constant background field $B_0$ in equation (\ref{eq:MeanFieldCriticalLambda}), so as to find the fastest growing MRI wavelength, thus disregarding the spatial variations. There is no theoretical justification for doing so, either in the linear or nonlinear regime, as locally the MRI will undoubtedly care about these variations. Note that the absolute value sign in the definition of $Q_z$ will retain the (spatially varying) turbulent fluctuations, which usually swamp the contribution from any (constant) background field. More specifically, the rapid changing of field polarity on small scales, as in current sheets \citep[abundant in MRI turbulence, see][]{sano2007,zhdankin17}, makes it difficult to attribute a well-defined linear MRI scale (see section \ref{sec:LinearTheory} later). 
In actual fact, a quasi-linear theory might be more justified in treating the fluctuations as a background \emph{turbulent diffusivity}, not a background mean field \citep[for e.g., see section 2.1 of][]{begelman2023}.

Meanwhile, the fluctuating local magnetic field is strongly time-dependent. In linear theory, the maximal growth rate is $s_\text{max} = (3/4) \Omega $, of a similar order to the temporal variability of the turbulent $B_z$ (e.g., see Fig. \ref{fig:TimeEvolutionQz}). Thus, there is no certainty that the linear theory (which presupposes a steady background field) applies, and that growth will be exponential. In fact, it has been shown elsewhere that linear instability can be delayed or suppressed in the presence of an unsteady background state \citetext{see \citealp{Levins1969} and \citealp{May1973}, in theoretical ecology, or \citealp{DeSwart1987}, in atmospheric modelling}. A notable example in the accretion disk context is thermal instability. \cite{Ross2017} showed that time-dependent MRI tubulence nullified any exponential growth, forcing the the unstable modes to behave as a biased random walk instead. Thus it seems unjustified to replace the time-independent background field by a time-averaged turbulent field so as to compute the quality factor.

\subsubsection{Energy injection in the non-linear saturated phase}\label{sec:Non-linearSaturation}

The definition of $Q_z$ relies on a linear analysis of the MRI. However, the MRI is only approximately linear during its rising phase. After a few tens of orbits, the MRI enters its highly non-linear saturated phase, in which plenty of modes with zero linear growth rate participate in the dynamics and achieve non-zero saturated amplitudes. A typical power spectrum of an ideal net-flux simulation consists of two zones:  an injection zone of roughly constant energy, and a turbulent cascade. Increasingly resolved simulations extend the cascade to increasingly higher wavenumbers \citep[e.g.,][]{Hawley1995,Simon2009}.  But, without explicit diffusion coefficients, the cascade can never be fully captured, and a high $Q_z$ guarantees, at best, that the injection scale, and its transfer of energy to nearby scales, is sufficiently resolved.

We argue that even this is unlikely. One might expect the MRI's linear theory to estimate the injection scales, which would then lie on  $\lambda>\lambda_\text{crit}(B_0)$ \citep[see, for e.g.,][]{kawazura24}, where $B_0$ is the true global \emph{background} magnetic field threading the domain, in contrast to that used in $Q_z$, i.e., the fluctuating \emph{local} field, $B_z$.
As we will see in section \ref{sec:NetFlux}, even in the net-flux case, $\lambda_\text{crit}(B_z) \gg \lambda_\text{crit}(B_0)$. It follows that the quality factor potentially overestimates the size of the injection scale, and consequently its resolvability: a high $Q_z$ does not guarantee the injection scale is resolved. 

Having said that, the true scale of maximum energy injection in the nonlinear saturated state is not fully constrained or understood; local net-flux simulations, in fact, reveal that the nonlinear injection scales fail to map cleanly on to the linear theory \citep{Lesur2011}. This could be due to non-axisymmetric structures dominating the energy injection. Specifically, maximum nonlinear injection occurs on scales larger than $\lambda_{\text{crit}}(B_0)$ (see fig. 3 of \cite{Walker2016}). We return to this issue in section \ref{sec:Discussions}.

\subsubsection{ZNF configurations and MRI dynamos}\label{sec:MRIdynamo}

Nonetheless, any weak background vertical flux (and its associated Alfv\'en length) remains a critical ingredient in the turbulence. If it is absent, such as in an initial ZNF configuration, the MRI may enter a  dynamo state that exhibits radically different properties to net-flux MRI turbulence. The underlying mechanisms sustaining this dynamo are fully nonlinear \citep[e.g.,][]{herault2011,riols2013,gressel2015,Held2022}, and clearly do not rely on a local linear stability analysis with a net-vertical flux. 

Moreover, the nonlinear local MRI dynamo is unconverged with resolution \citep{Fromang2007,Ryan2017}.\footnote{Though note the exception of a sufficiently tall, thin unstratified box \citep{Shi2016}.} To obtain convergence, explicit diffusion coefficients must be included \citep{Fromang2007nonideal}, which would then render quality factors needless, as long as there were enough grid cells to resolve the viscous and resistive scales (admittedly, something difficult to estimate a priori especially with the Hall effect and ambiopolar diffusion). We interrogate the fate of the quality factor in ideal ZNF boxes in more detail in section \ref{sec:Simulation}.

\subsubsection{Generalising from the local to the global}\label{sec:locglob}

As discussed in section \ref{sec:Qphi}, in the context of linear theory, it is not straightforward to take shearing box results and apply them to global disk models \citep[though see][]{latter2015}. In particular, what does the non-convergence of the local ZNF MRI mean (if anything) for global MRI turbulence?\footnote{In fact, it is uncertain if ZNF vertically stratified simulations remain unconverged when the radial domain size is sufficiently extended. The example of thermal instability in the radiation-dominated regime presents a cautionary tale \citep{jiang2013}.} 


It might be argued that, at a given time, if a disk annulus of sufficient width is threaded by a net-vertical flux, exhibiting little radial variation, then the MRI's behaviour in that annulus might approximate net-flux shearing boxes \citep[cf.,][]{sorathia2010,Sorathia2012}. On the other hand, if the vertical flux oscillates rapidly with radius, then the annulus might behave more like a ZNF box. A key question would then be: what is the critical lengthscale of radial variation above which portions of the disk behave as if net flux ? This approach to the problem will, of course, be complicated by the slow evolution of the vertical flux, which might bunch up into rings or pile up at one of the numerical boundaries, evacuating much of the domain \citep[e.g.][]{johansen2009,Bai2014}. It also critically assumes that the MRI in global models behaves like a sum of radially adjacent manifestations of the local MRI. But perhaps the MRI works across a large range of radii concurrently, and is thus a fully global process \citep[see recent work by][]{Jacquemin2024}. If this is indeed the case, the utility of a measure derived from purely local processes, i.e. the quality factor, is likely limited.

\section{Large quality factors in unconverged zero-net-flux simulations}\label{sec:Simulation}

In their seminal work, \citetalias{Fromang2007} showed that, in ideal MHD, unstratified ZNF simulations are unconverged. Turbulence scales (e.g. the correlation length) are closely tied to the grid scale, which also sets the various stresses. In this section we reproduce their simulations using the new GPU-accelerated code \texttt{IDEFIX} \citep{Lesur2023}\footnote{\texttt{IDEFIX} is freely available at \href{https://github.com/idefix-code/idefix}{https://github.com/idefix-code/idefix}}. Our aim is to follow the evolution of the quality factors with resolution, and to try to understand their behaviour.

The \texttt{IDEFIX} code is a conservative finite-volume Godunov code, with a design close to that of the \texttt{PLUTO} code \citep{Mignone2007}. It relies on the \texttt{KOKKOS} library \citep{Kokkos} for GPU parallelization. Its high energy efficiency and performance (see section 5 of \cite{Lesur2023}) allowed us to run the 12 simulations presented in this paper in only around 3000 GPU hours.

\subsection{Ideal MHD equations}\label{sec:IdealMHDequations}

In this paper we use the shearing box approximation. To this end, we consider a cartesian system of coordinates $(x,y,z)$ with unit vectors $(\bm{e}_x,\bm{e}_y,\bm{e}_z)$. Those vectors are oriented in the radial, azimuthal and vertical direction respectively. The frame rotates with rotation vector $\bm{\Omega} = \Omega \bm{e}_z$, where $\Omega$ is the angular velocity of a particle located at the centre of the box and origin of the frame. In this frame we assume ideal MHD, and the equations are:

\begin{align}
&\frac{\partial \rho}{\partial t} + \nabla \cdot \left( \rho \bm{u} \right) = 0, \label{eq:MassConservation} \\
&\frac{\partial {\bm{u}}}{\partial t} + \left( \bm{u} \cdot \nabla \right) \bm{u} = - \frac{\nabla P}{\rho} - 2 \bm{\Omega} \times \bm{u} + \frac{1}{4 \pi\rho} \left( \nabla \times \bm{B} \right)\times \bm{B} + 3 \Omega^2 \bm{e}_x, \label{eq:Momentum} \\
&\frac{\partial \bm B}{\partial t} + \nabla  \times \left(\bm B \times \bm u \right)  = \bm{0}. \label{eq:Induction}
\end{align}

\noindent{}where $\rho$ is the density, $\bm{u}$ the velocity, $P$ the thermal pressure and $\bm{B}$ the magnetic field. 

For the sake of simplicity and comparison to other numerical works, we use an isothermal equation of state $P = \rho c_0^2$, where $c_0$ is the constant sound speed. It is used to define the disk scale height $H = c_0 / \Omega$. All runs but those presented in section \ref{sec:Stratified} use the `unstratified' box: there is no vertical structure or vertical component to the tidal potential.

\subsection{Numerical setup}\label{sec:NumericalSetup}

Similar to \citetalias{Fromang2007}, our numerical domains are of size $(L_x,L_y,L_z)=(H,\pi H, H)$, and we run three simulations (STD64, STD128 and STD256) of respective resolutions $(N_x,N_y,N_z) = (64,100,64)$, $(128,200,128)$ and $(256,400,256)$.

Thermal pressure and density are initially uniform. As those simulations are in the zero-net-flux configuration, the initial magnetic field is initially purely vertical with the profile:
\begin{equation} \label{eq:ZeroNetFlux}
    B_z = B_0 \sin{(2 \pi x /H)},
\end{equation}
\noindent{}where $B_0$ is chosen such that initially, the domain averaged plasma $\beta$, i.e. the ratio of thermal and magnetic pressures, is 400.

The initial velocity has two components. One is due to the Keplerian shear, $-(3 \Omega x /2 ) \bm{e}_y$. The other is random velocity variations of small amplitude ($10^{-5} c_0$) in all directions, to initiate the instability.

The boundary conditions are periodic in $y$ and $z$ and shear-periodic in $x$ \citep{Hawley1995}. We use the HLLD Riemann solver  \citep{Miyoshi2005}. For time-stepping we use a second order Runge-Kutta method with a Courant-Friedrichs-Lewy (CFL) constant $C=0.8$. The solenoidal condition is enforced via a constrained transport method \citep{Evans1988}, the $\mathcal{E}_c$ scheme of \cite{Gardiner2008}.

\subsection{Averaging procedures and diagnostics}

To quantify the turbulence in our system, we need to define a set of averages. The space-average of any quantity $f$ over the whole computational domain is defined as

\begin{equation}
    \left< f \right>_{x,y,z} = \frac{1}{V} \int\limits^{L_x/2}_{-L_x/2} \int\limits^{L_y/2}_{-L_y/2} \int\limits^{L_z/2}_{-L_z/2} f dx dy dz ,
\end{equation}

\noindent{}where $V = L_x L_y L_z$ is the total volume of the domain.

An average in space and time is then defined as

\begin{equation}
    \left< f \right>_{t,x,y,z} = \frac{1}{t_2 - t_1} \int\limits^{t_2}_{t_1} \left< f \right>_{x,y,z} dt ,
\end{equation}

\noindent{}where $t_2 = t_f$, the final integration time of the simulation, and $t_2-t_1 = 100$ orbits, except for sections \ref{sec:NetFlux} and \ref{sec:Stratified} where $t_2-t_1 = 40$ orbits.

To study the vertical profile of the turbulence, we introduce averages on $(x,y)$ and on $(t,x,y)$, defined respectively as

\begin{equation}
    \left< f \right>_{x,y} = \frac{1}{L_x L_y} \int\limits^{L_x/2}_{-L_x/2} \int\limits^{L_y/2}_{-L_y/2} f dx dy ,
\end{equation}

\begin{equation}
    \left< f \right>_{t,x,y} = \frac{1}{t_2 - t_1} \int\limits^{t_2}_{t_1} \left< f \right>_{x,y} dt .
\end{equation}

The Reynolds, Maxwell and total stresses are averaged in space and time and normalized by the initial thermal pressure $P_0$. They are defined respectively as:

\begin{align}
& \alpha_R = \frac{1}{P_0} \biggl \langle \rho u_x \left( u_y - 3/2 \Omega x \right) \biggr \rangle_{{t,x,y,z}} \hspace{-0.82cm},\\
& \alpha_M = \frac{1}{P_0} \left< - \frac{B_x B_y}{4 \pi} \right>_{t,x,y,z}  \hspace{-0.87cm},\\
& \alpha = \alpha_R + \alpha_M.
\end{align}

Several correlation lengths could be defined, which all measure the characteristic length scales of the turbulence. Following \cite{Lesur2007} and \citetalias{Fromang2007}, we adopt the following representative vertical definition, which is sufficient for our purposes:

\begin{equation}\label{eq:CorrelationLength}
    \mathscr{L}_z(B_y) = \left< \frac{\left( \int  B_y(y=0) dz \right)^2}{\int B_y(y=0)^2 dz}\right>_{t,x} \hspace{-0.4cm}.
\end{equation}

In the following, units are chosen so that $2 \pi / \Omega=1$, $\rho_0=1$ for density, and $H=1$.


\subsection{The simulations are unconverged}

The three simulations are listed on Table \ref{tab:UnstratifiedRuns} with their main saturation properties. After a growing phase of 10 to 30 orbits, they all reach a saturated state. All simulations were run over hundreds of orbits, to provide averages on the state of the saturated turbulence.

The stresses measured in our simulations follow the same trend with resolution as in \citetalias{Fromang2007}. The decrease is slightly slower than their observed $\alpha \propto 1/N$, and the Maxwell stress is twice as large. In such unresolved simulations the numerical scheme naturally plays a significant role in setting the level of the turbulence. The slower decrease might be due to our higher order reconstruction scheme \citep{Bodo2011}. Meanwhile, the larger stresses arise from using a finite-volume code (\texttt{IDEFIX}); \citetalias{Fromang2007} used the more diffusive finite-difference code \texttt{ZEUS} \citep{HawleyStone1995}. This is in agreement with the simulations of \cite{Simon2009}, performed on the finite-volume code \texttt{ATHENA} \citep{Stone2008}. Their simulation SZ128 is akin to our simulation STD128, but with stresses about 70\% larger than STD128.

\begin{figure}
	\includegraphics[clip,trim=0 10 0 5, width=\columnwidth]{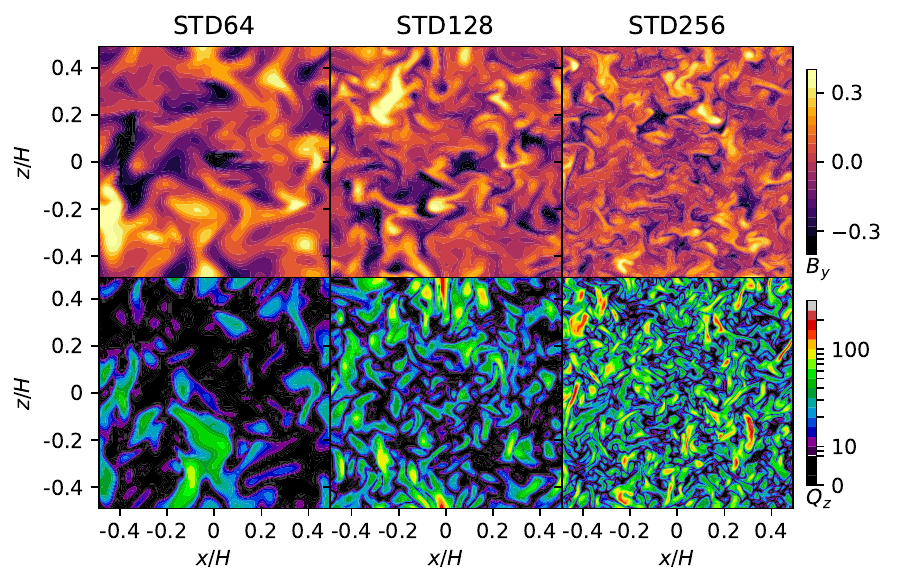}
    \caption{Distribution of $B_y$ (first row) and $Q_z$ (second row) in the $(x,z)$ plane $y=0$ for the last outputs of simulations STD64, STD128 and STD256. In the bottom row all cells with $Q_z < 8$ are shown in black. From left to right, the increase in resolution leads to the appearance of smaller structures.}
    \label{fig:CoupeBy}
\end{figure}

The top row of Fig. \ref{fig:CoupeBy} is analogous to fig. 2 of \citetalias{Fromang2007}. It shows cuts at $y=0$ at the last timestep of all three simulations. The background colour describes the toroidal magnetic field $B_y$. We can see that, from left to right, as the resolution increases the scale of the turbulence diminishes. This can be quantified via the correlation length $\mathscr{L}_z$. By fitting it to the grid cell size, we obtain $\mathscr{L}_z \propto \delta z ^{0.31}$, where $\delta z$ is the cell size in the vertical direction. This indicates non-convergence, in line with the behaviour of the stresses.

\begin{table}
	\centering
	\caption{Simulations STD64, STD128 and STD256, of respective resolutions $(N_x,N_y,N_z) = (64,100,64)$, $(128,200,128)$ and $(256,400,256)$. $t_f$ (in orbits) is the final timestep of the run; $\alpha_R$, $\alpha_M$ and $\alpha = \alpha_R + \alpha_M$ are the averaged Reynolds, Maxwell and total stresses; $\mathscr{L}_z$ (in units of box size $H=L_z$) is the correlation length (equation \ref{eq:CorrelationLength}); $\mathcal{R}_z$ is the vertical resolvability (equation \ref{eq:Resolvability}). All quantities but $\mathcal{R}_z$ are space-averaged over the whole box and time-averaged over the last 100 orbits.}
	\label{tab:UnstratifiedRuns}
	\begin{tabular}{ c c c c c c c } 
		\hline
		Run & $t_f$ & $10^4 \alpha_R$ & $10^4  \alpha_M$ & $ 10^4 \alpha $ & $10^2 \mathscr{L}_z $ & $\mathcal{R}_z$\\
		\hline
		STD64 & 726 & $9.9$ & $71$ & $81$ & 8.3 & 0.49\\
		STD128 & 635 & $6.0$ & $44$ & $50$ & 6.6 & 0.71\\
		STD256 & 385 & $3.4$ & $26$& $30$ & 5.4 & 0.80\\
		\hline
	\end{tabular}
\end{table}

\begin{figure}
	\includegraphics[clip,trim=0 8 0 5,width=\columnwidth]{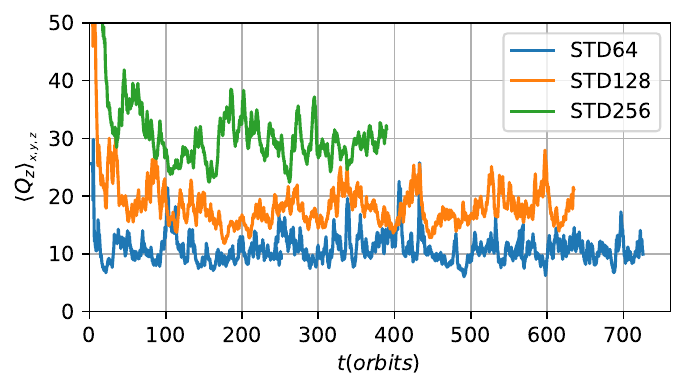}
    \caption{Evolution of the space-averaged $\left< Q_z \right>_{x,y,z}$ over time.}
    \label{fig:TimeEvolutionQz}
\end{figure}

\subsection{The simulations exhibit large quality factors}\label{sec:LargeQualityFactors}

We now compute the quality factor $Q_z$, in order to see its evolution with resolution. The second row of Fig. \ref{fig:CoupeBy} shows cuts of $Q_z$ at $y=0$, for the last snapshot of each simulation. $Q_z$ varies on a lengthscale similar to that of the turbulence, and the maximum quality factor increases with resolution. The statistics of $Q_z$ in the simulations are illustrated in appendix \ref{appendix:Histograms}. Fig. \ref{fig:TimeEvolutionQz} shows timeseries of $Q_z$, space-averaged over the whole simulation box. We see that after a few tens of orbits, the quality factor fluctuates around a mean saturated value.

The mean quality factor (averaged over the last $100$ orbits) is plotted against resolution in Fig. \ref{fig:ResolutionQz}. We see a clear increase with resolution, with $Q_z \simeq 30$ for simulation STD256. This value is well over $Q_z \simeq 10 - 15$ usually taken as a threshold for convergence, even though the simulation is not numerically convergent. This proves that at least for this setup, the quality factor fails as a metric of numerical convergence.

The last column of Table \ref{tab:UnstratifiedRuns} shows the vertical resolvability $\mathcal{R}_z$ (equation \ref{eq:Resolvability}) for each simulation at its final timestep. For simulations STD128 and STD256 it is over $0.7$, the common threshold for simulation convergence. This shows that, at least for this setup, $\mathcal{R}_z$ is not a good convergence indicator either.

\subsection{Lengths other than cell size influence the turbulence}\label{sec:LengthsOtherThanCellSize}

Let us try to understand why the quality factor does not work in this case. In a simplified view, its use rests on the idea that the Alfvén length (or any length characteristic of the turbulence), should behave in one of the following two ways:

\begin{itemize}
    \item If a simulation is resolved, then the Alfvén length is constant with respect to resolution, and $Q_z$ should increase linearly with $\delta z$.
    \item If a simulation is unresolved, then the Alfvén length is proportional to $\delta z$ and $Q_z$ is then independent of $\delta z$.
\end{itemize}

\begin{figure}
	\includegraphics[clip,trim=0 8 0 5,width=\columnwidth]{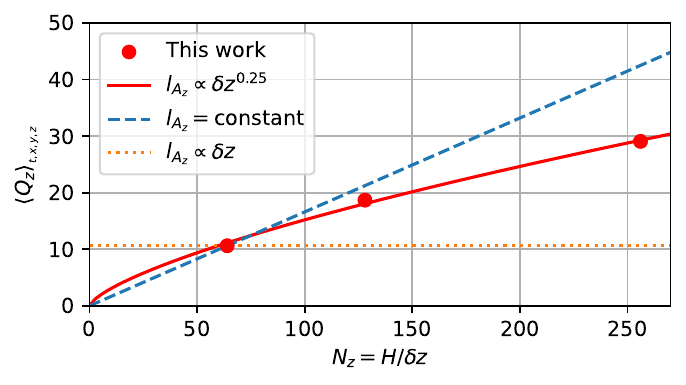}
    \caption{Points: evolution of $ \left< Q_z \right>_{t,x,y,z} $ with resolution.
    Red line: best fit of those points ($l_{A_z} \propto \delta z^{0.25}$). Blue line: hypothetical resolved case ($l_{A_z}$=constant). Orange line: hypothetical non-resolved case ($l_{A_z} \propto \delta z$).}
    \label{fig:ResolutionQz}
\end{figure}

In Fig. \ref{fig:ResolutionQz} we represent how these hypotheticals would behave, respectively in blue and orange. We can see that the red points corresponding to our simulations sit between the two, with $l_{A_z} \propto \delta z ^{0.25}$. Thus as the Alfvén length scales with the grid scale but is not directly proportional to it, the quality factor keeps increasing with resolution even though the simulations are unresolved.

But why is the Alfvén length not directly proportional to the grid cell size ? As there is no physical dissipation in ideal MHD, one might think the grid scale is the only relevant length scale of the problem for sufficently high resolutions. On the contrary, \cite{Shi2016} showed that the vertical size of the box also controls the state of the turbulence. Increasing the aspect ratio $L_z / L_x$, they show an increase of stresses until $L_z / L_x \sim 2.5$, beyond which they uncover a new dynamo regime.

\begin{figure}
	\includegraphics[clip,trim=0 8 0 5,width=\columnwidth]{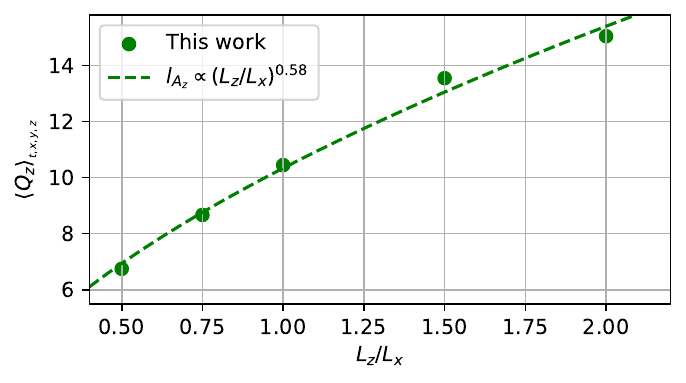}
    \caption{Points: evolution of $ \left< Q_z \right>_{t,x,y,z} $ with vertical aspect ratio. Line: best fit of those points ($l_{A_z} \propto (L_z / L_x)^{0.58}$). All simulations have the same resolution as STD64: $(N_x,N_y,N_z) = \left(  64,100,64(L_z/L_x)\right)$.}
    \label{fig:AspectQz}
\end{figure}

To see how this would affect the quality factor, we performed simulations varying the aspect ratio, all other things being equal. We started from simulation STD64, with $L_z = H$, and performed simulations with $L_z \in \{ 0.5 H; 0.75 H; 1.5 H; 2 H \}$, keeping $L_x = H$, $L_y = \pi H$ and the grid cell sizes unchanged. On Fig. \ref{fig:AspectQz} we show how $Q_z$ behaves with the aspect ratio. It increases, consistently with the increase of $\alpha$ with aspect ratio seen in \cite{Shi2016}. We can fit the monomial $l_{A_z} \propto (L_z/L_x)^{0.58}$ on those points.

Consequently, in unresolved simulations the turbulent scale is not directly proportional to $\delta z$, because there is at least one other length critical to the problem, the vertical size of the box. However, as $l_{A_z} \propto \delta z^{0.25} L_z^{0.58}$ and $ 0.25+0.58 = 0.83 < 1$, on dimensional grounds there are other lengths at play. The radial box size $L_x$ should have a weaker although noticeable effect \citep{Shi2016,Held2022}. The scale height $H = c_s / \Omega$ has been shown to influence the stresses of MRI turbulence \citep{Pessah2007}. Thus we might expect the Alfvén length to depend on the following scales:

\begin{equation}\label{eq:MixedScaling}
    l_{A_z} \propto \delta z^a L_z^b L_x^c H^d
\end{equation}

\noindent{}where $a$, $b$, $c$ and $d$ are positive exponents such that $a+b+c+d = 1$, and assuming cubic cells. Other length scales characteristic of the turbulence such as the correlation length should exhibit similar behaviours. The exponents will depend on the numerical scheme, shown to modify the scalings with resolution \citep{Bodo2011}.

As a summary, in the ideal zero-net-flux configuration, there are scales unrelated to the grid scale size or to a physical dissipation mechanism that contribute to setting the turbulent scale. Moreover, the characteristic turbulence scale is not directly proportional to the grid cell size, even for unresolved simulations, but follows a `mixed' scaling (equation \ref{eq:MixedScaling}). This makes a critical difference to the behaviour of $Q_z$: for sufficiently small $\delta z$ the quality factor will be large enough to indicate convergence, but the simulation is in fact unconverged. Note that the quality factors $Q_x$ and $Q_y$ exhibit a similar behaviour, illustrated in appendix \ref{appendix:QxQy}.

\subsection{The net-flux case}\label{sec:NetFlux}

In the above, we studied the quality factor in the zero-net-flux case. Here we will compare it to the net-flux case. For this purpose, we ran two additional simulations. In those, the same ideal MHD equations are solved (section \ref{sec:IdealMHDequations}) and we use an almost identical setup (section \ref{sec:NumericalSetup}). The only change is the initial magnetic field (equation \ref{eq:ZeroNetFlux}). In the net-flux case it is still purely vertical, but now constant in the whole box: $B_z = B_0/2$. The factor $1/2$ is there to set $\left< \beta (t=0) \right>_{x,y,z} = 400$, as in the zero-net-flux simulations. Those two simulations are named NF64 and NF128. Their resolutions and box sizes are identical to STD64 and STD128, respectively.

Both simulations are listed on Table \ref{tab:NetFluxRuns}. They quickly reach a saturated state, after about 5 orbits. As expected, the stresses are much larger than in the zero-net-flux case, and appear independent of resolution. The indicators also predict convergence: for both simulations, $Q_z$ is much greater than 15, and $\mathcal{R}_z$ is very close to unity. This seems a good endorsement of the quality factor: the simulations appear converged, and $Q_z$ predicts convergence.

However, we argue that its use is also problematic in this case. It is defined as $Q_z = \sqrt{3} \lambda_{\text{crit}}(B_z) / \delta z$ (equation \ref{eq:DefinitionQz}), where $\lambda_{\text{crit}}$ is derived from linear theory (equation \ref{eq:MeanFieldCriticalLambda}). In this approach, $\lambda_{\text{crit}}$ is thought to be the wavelength of the smallest possible MRI mode. In our simulations, this gives $\lambda_{\text{crit}}(B_z) = 0.9 L_z$ for NF64 and $1.1 L_z$ for NF128, meaning that for NF128 the critical wavelength is larger than the box size\footnote{Magnetically arrested disks often have $\lambda_{\text{crit}}>H$, which wrongly lead to the belief that MRI is quenched (see \cite{Begelman2022}, section 2.4).}. Thus the assumptions behind the quality factor would mean that there is no MRI mode that could grow within the boundaries of the numerical domain. This absurdity stems from the definition of the quality factor mentioned in section \ref{sec:Non-linearSaturation}: it uses the absolute value of the local magnetic field $\lvert B_z \rvert$, while it is the background magnetic field that appears in the critical wavelength, here $B_0/2$. We get $ \left< \lvert B_z \rvert \right>_{t,x,y,z} / (B_0 / 2) = 4.5$ for NF64 and 5.7 for NF128, as the turbulent field dwarfs the background field (see also appendix \ref{appendix:NetFluxBz}). Thus using the local magnetic field $B_z$ rather than the background field $B_0 / 2$ in $Q_z$ leads to a likely overestimation of the quality of the simulation.

\begin{table}
	\centering
	\caption{Simulations NF64 and NF128, of respective resolutions $(N_x,N_y,N_z) = (64,100,64)$ and $(128,200,128)$. $t_f$ (in orbits) is the final timestep of the run; $\alpha_R$, $\alpha_M$ and $\alpha = \alpha_R + \alpha_M$ are the averaged Reynolds, Maxwell and total stresses; $Q_z$ is the quality factor; $\mathscr{L}_z$ (in units of box size $H=L_z$) is the correlation length (equation \ref{eq:CorrelationLength}); $\mathcal{R}_z$ is the vertical resolvability (equation \ref{eq:Resolvability}). All quantities but $\mathcal{R}_z$ are space-averaged over the whole box and time-averaged over the last 40 orbits.}
	\label{tab:NetFluxRuns}
	\begin{tabular}{ c c c c c c c c } 
		\hline
		Run & $t_f$ & $10^2 \alpha_R$ & $10^2  \alpha_M$ & $ 10^2 \alpha $ & $Q_z$ & $10^2 \mathscr{L}_z $ & $\mathcal{R}_z$\\
		\hline
		NF64 & 91 & 5.6 & 50 & 55 & 97 & 34 & 0.96\\
		NF128 & 119 & 4.9 & 48 & 53 & 242 & 91 & 0.99\\
		\hline
	\end{tabular}
\end{table}

\subsection{The zero-net-flux stratified case}\label{sec:Stratified}

\begin{figure}
	\includegraphics[clip,trim=0 8 0 6,width=\columnwidth]{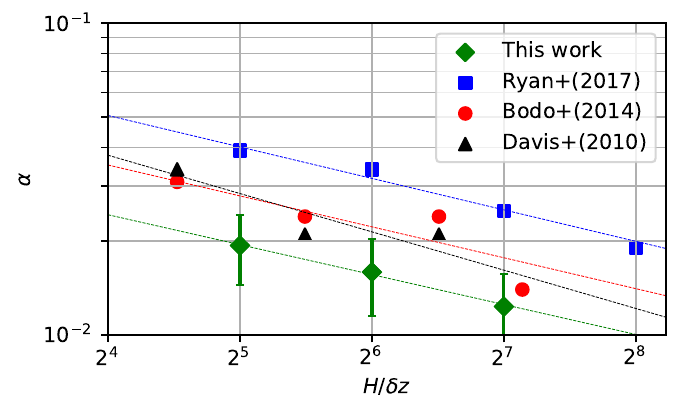}
    \caption{Time and space-averaged stress $\alpha$ as a function of resolution for ZNF stratified simulations. Dashed lines are fits of $\alpha \propto (H/ \delta z)^{-a}$ to each dataset, where $a$ is 0.32 (this work), 0.34 \citep{Ryan2017}, 0.33 \citep{Bodo2014} and 0.41 \citep{Davis2010}. Error bars are standard deviations with time of the space-averaged stresses, assumed to be stationary processes.}
    \label{fig:AlphaResolution}
\end{figure}

\begin{figure}
	\includegraphics[clip,trim=0 8 0 6,width=\columnwidth]{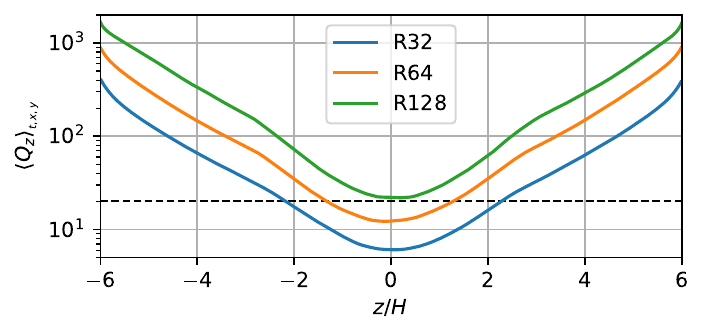}
    \caption{Vertical profiles of $ \left< Q_z \right>_{t,x,y}$ for simulations R32, R64 and R128. For R128 it is larger than 20 at all altitudes.}
    \label{fig:StratifiedQzProfiles}
\end{figure}

Finally, we consider the case where vertical stratification is no longer neglected. It is introduced via an additional gravity term $ - (\rho \Omega^2 z )\bm{e}_z$ in the right-hand side of the momentum equation (equation \ref{eq:Momentum}). For $\bm{B} = \bm{0}$, this new momentum equation and equation (\ref{eq:MassConservation}) admit an equilibrium state $\rho = \rho_0 \exp({-z^2/(2 H^2))}$ and $\bm{u} = - (3 \Omega x /2 ) \bm{e}_y$. We take it as the initial conditions of the simulations, adding small velocity perturbations to initiate the instability.

We reproduce the simulations R32, R64 and R128 of \cite{Ryan2017}. The box sizes are $(L_x, L_y, L_z) = (3H, 4H, 12H)$ and the resolutions $(N_x, N_y, N_z) = (96, 64, 384)$, $(192, 128, 768)$ and $(384, 256, 1536)$ respectively. The boundary conditions are shear-periodic in $y$, periodic in $x$ and outflow in $z$: we enforce $ \partial \bm{B} / \partial z = \partial \bm{u} / \partial z = \bm{0} $, and $ c_0^2(\partial \rho /\partial z) = - \rho \Omega^2 z$. The only differences with their setup are the use of the \texttt{IDEFIX} code rather than \texttt{RAMSES-GPU} \citep{Kestener2017}, and the initial magnetic field configuration. Our initial field is initially purely vertical with a profile $B_z = B_1 \sin{(2 \pi x / H)}$, where $B_1$ is chosen such that $\left< \beta (t=0,z=0) \right>_{x,y} = 50$. We verified that a field initially purely toroidal leads to the same results.

Such zero-net-flux stratified simulations are known to be unconverged since \cite{Bodo2014}. Our Fig. \ref{fig:AlphaResolution} is analogous to fig. 9 of \cite{Ryan2017}. It shows the decrease of the total stress with resolution, as $\alpha \propto (H/ \delta z)^{-1/3}$. Note that differences in units with \cite{Davis2010} and \cite{Bodo2014} required conversions (see section 4 of \cite{Ryan2017}).

These unconverged simulations yield large quality factors. In Fig. \ref{fig:StratifiedQzProfiles} we present the vertical profiles of $ \left< Q_z \right>_{t,x,y}$. We see the expected rise of $Q_z$ from the midplane to the edges of the box. Because of that, \cite{Hawley2011} only considered $Q_z$ for $\lvert z \rvert \leq 0.5 H$. We see in Fig. \ref{fig:StratifiedQzProfiles} that even then, R128's quality factor is much larger than 15, the usual threshold for simulation convergence. Note that the most resolved simulation of \cite{Davis2010} reached $Q_z \simeq 18$ in our units in the midplane region (see table 1 of \cite{Hawley2011}). The quality factors $Q_x$ and $Q_y$ are also very large (see appendix \ref{appendix:StratifiedQxQy}).

As a summary, stratified simulations are also unconverged, as their stresses and correlation lengths \citep{Ryan2017} depend on the grid cell size. This remains the case at higher resolutions, where we find that the midplane quality factor is larger that the usual threshold required for convergence. In conclusion, even in stratified simulations, the quality factor indicates convergence even when the MRI is clearly not converged, and thus fails as a diagnostic.

\subsection{The strongly magnetized zero-net-flux case}\label{sec:StronglyMagnetized}

In this paper, we focus on the weakly-magnetized case ($\beta \gg 1$). \cite{Squire2025} recently published ideal zero-net-flux simulations that reveal a different dynamo state, in which the midplane is strongly magnetized ($\beta \lesssim 1$). Their simulation $\beta 0.01$ of resolution $336^3$ yields $\alpha \simeq 0.61$, and $Q_z >20$ in the midplane. Yet when restarted at double the resolution ($672^3$, simulation $\beta 0.01$-hr), all other things being equal, the dynamo yields $\alpha \simeq 0.46 $, indicating non-convergence with respect to the normalised stress. Though we have only two data points, and further tests are clearly warranted, it would appear that the quality factor is not a reliable indicator of convergence for strongly magnetized simulations either.

\section{Linear theory of the MRI in the presence of a magnetic null}\label{sec:LinearTheory}


In this section, we approach the quality factor from a different angle to that of section \ref{sec:Simulation}, and instead interrogate its underlying linear theory. As explained in section \ref{sec:TheoreticalBackground}, $Q_z$ takes the spatially varying local field at each grid point from a turbulent simulation and conflates it with the background net field in the classical MRI linear theory. This process supplies a local characteristic MRI lengthscale that can be spatially averaged and then compared with $\delta z$. Section \ref{sec:SpaceVaryingMagneticFields} discusses some general objections to this approach, in particular the doubt that MRI modes will remain oblivious to neighbouring variations in the magnetic field, especially if they are on a similar wavelength and if they involve changes in field polarity. 
Here we undertake a set of idealised linear calculations that flesh out some of these points. Instead of calculating MRI modes at each grid point, using the local $B_z$, with each grid point completely independent of each other, we calculate the modes self-consistently in a small radial region in which $B_z$ varies radially. Our analysis shows that, in the presence of a magnetic null, there is no secure or single MRI lengthscale, in contrast to the classical linear theory. In fact, we uncover growth on arbitrarily small scales, which confounds the evaluation of a local quality factor in this case.



\subsection{Perturbation equations}

To the ideal equations described in section \ref{sec:IdealMHDequations} we add the hypothesis of incompressibility. Equations (\ref{eq:MassConservation}) and (\ref{eq:Induction}) then become:
\begin{align}
    & \nabla \cdot \bm{u} = \bm{0}, \\
    & \frac{\partial \bm{B}}{\partial t} + \left( \bm{u} \cdot \nabla \right) \bm{B} = \left( \bm{B} \cdot \nabla \right) \bm{u}. \label{eq:IncompressibleInduction}
\end{align}
Equation (\ref{eq:Momentum}) is unchanged and we do not need an equation of state. We consider an equilibrium state with velocity $- ( 3 \Omega x /2 ) \bm{e}_y$, constant thermal pressure $P_0$ and radially varying magnetic field $B_0 f(x/\ell) \bm{e}_z$ where $B_0$ is a constant, $\ell$ a characteristic length scale and $f$ a dimensionless function. We only ask that $f$ has at least one zero.

We consider small axisymmetric perturbations $\bm{u}'=\bm{v}$, $P'$, $\bm{B}'= B_0 \bm{b}$ $\propto e^{i k_z z + s t}$. Using equations (\ref{eq:IncompressibleInduction}), (\ref{eq:Momentum}) and $\nabla \cdot \bm{u} = \nabla \cdot \bm{B} = 0$, we obtain the following linear equations:
\begin{align}
s v_x &= 2\Omega v_y - \partial_x h + i k_z V_{A}^2 f b_x, \label{eq:x-momentum} \\
s v_y & = -\frac{1}{2}\Omega v_x + i k_z V_{A}^2 f b_y, \label{eq:y-momentum} \\
s v_z & = - i k_z h + i k_z V_{A}^2 f b_z + V_{A}^2 (\partial_x f) b_x, \label{eq:z-momentum} \\
s b_x &= i k_z f v_x, \label{eq:x-induction} \\
s b_y &= i k_z f v_y - \frac{3}{2}\Omega b_x, \label{eq:y-induction} \\
s b_z &= i k_z f v_z - \partial_x f v_x, \label{eq:z-induction} \\
\partial_x v_x &= - i k_z v_z, \label{eq:incompressibility} \\
\partial_x b_x &= - i k_z b_z, \label{eq:solenoidal}
\end{align}
\noindent{}where $h=P'/\rho$ and $V_{A}^2 = B_0^2 / (4 \pi \rho)$ is the squared Alfvén speed.

If we assume boundary conditions, a magnetic field profile $f$ and values for $\Omega$, $V_A$, $\ell$ and $k_z$, we can solve equations (\ref{eq:x-momentum}) to (\ref{eq:solenoidal}) for the growth rate $s$.
There are two key dimensionless input parameters: a scaled vertical wavenumber, $K= k_z l_A$, and a scaled Alfvén length, $\epsilon = l_A / \ell$, where $l_A=V_A/\Omega$. The weak field limit might be thought of as the regime $ 0 < \epsilon \ll 1$. 

\subsection{Numerical solutions: Positive growth rates at inifinitely small scales}\label{sec:PositiveGrowthRatesInfinitelySmallScales} 

\begin{figure}
	\includegraphics[clip,trim=0 8 0 6,width=\columnwidth]{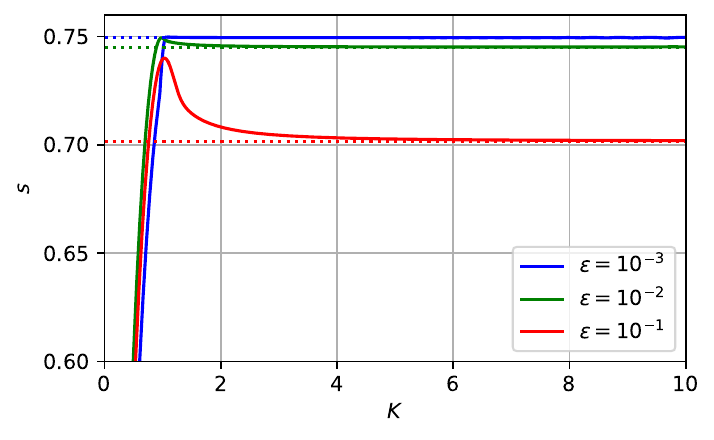}
    \caption{Solutions to the linearised equations for $\epsilon \in \{10^{-3};10^{-2};10^{-1} \}$. Solid lines correspond to the maximal growth rates retrieved by the pseudo-spectral method solving equations (\ref{eq:x-momentum}) to (\ref{eq:solenoidal}). Dotted lines are their asymptotic estimates (equation \ref{eq:FastestGrowingMode} for $n=0$). Each color represents a different $\epsilon$. For small values of $\epsilon$, the growth rates reach a plateau of height close to $(3/4) \Omega$, the maximal growth rate in the usual net flux MRI theory.}
    \label{fig:GrowthRateWavenumber}
\end{figure}

The equations (\ref{eq:x-momentum}) to (\ref{eq:solenoidal}) are solved numerically, assuming $f = \sin{(x/\ell)}$ and $2 \pi $ periodicity in $x$\footnote{Note that this is a common initial condition in ZNF simulations.}. Following \cite{Boyd1996}, we use a pseudo-spectral method to reframe the problem as an algebraic eigenvalue problem, which we then solve via a partial eigensolving algorithm \citep[see, e.g.,][]{GolubVanLoan1996}. At each wavenumber $K$ we look for six eigenvalues. The largest eigenvalues are plotted as solid lines on Fig. \ref{fig:GrowthRateWavenumber}, for several values of $\epsilon$. We find that at large wavenumbers the growth rates hits plateaus: there is no decline of $s$ at large $K$ (small scales). We also show that as $\epsilon$ decreases, the plateau becomes closer to $(3/4) \Omega$, the maximal growth rate in the classical linear MRI theory (with a constant magnetic field).





\subsection{Asymptotic analysis}

\subsubsection{Reduction to Schrödinger form}

We will now explore the problem analytically, in order to find its dispersion relation in the limit of small-scale modes and weak fields. Using equations (\ref{eq:x-momentum}) to (\ref{eq:solenoidal}) we obtain the following differential equation for $v_x$:
\begin{equation} \label{eq:ODE1}
 \frac{d}{dx}\left(s_A^2\frac{d v_x}{dx}  \right) - \frac{k_z^2}{s_A^2}D(s, V_{A} k_z f)v_x =0,
 \end{equation}
where $s_A^2 = s^2 +V_{A}^2 k_z^2 f^2$ and $D$ is the usual dispersion relation (equation \ref{eq:DispersionRelationMeanField}) but with the inclusion of the function $f$. We keep $f$ generic for now, only asking that it has at least one zero.

To study its asymptotic behaviour, we put equation (\ref{eq:ODE1}) in Schrödinger form, with $\Psi = s_A v_x$. It becomes:
 \begin{equation} \label{eq:ODE2}
 \frac{d^2\Psi}{dx^2} - \frac{k_z^2}{s_A^4}D(s,V_{A} k_z f) \Psi + \frac{1}{2 s_A^4}\left[\frac{1}{2}\left(\frac{d s_A^2}{dx}\right)^2-s_A^2 \frac{d^2 s_A^2}{dx^2}   \right]\Psi=0.
 \end{equation}
The relative size of the third to the second term is $\sim K^2/\epsilon^2$. Thus either in the weak field limit ($0<\epsilon\ll 1$) or the short wavelength limit $(k_z \gg l_A)$, we may safely drop the third term. Next we adopt units so that $\Omega=1$ and $\ell=1$ and obtain
\begin{align} \label{eq:ODE4}
    \frac{d^2\Psi}{dx^2} - \frac{K^2}{\epsilon^2} V(x) \Psi =0,
\end{align}
in which the potential $V$ is defined by
\begin{equation}\label{eq:PotentialV1}
V= \frac{D(s, Kf)}{s_A^4} = \frac{s_A^4-3 s_A^2 +4 s^2}{s_A^4}.
\end{equation}

\subsubsection{Turning points and mode localisation}
The potential exhibits turning points wherever $V=0$, and thus the domain may be subdivided into regions where $V<0$ (`wave zones') and $V>0$ (`evanescent zones'). We expect our solutions to be localised in the former. 
Note that the turning points occur where $D=0$, i.e. when the usual dispersion relation is satisfied. The turning points are the roots of the following equation:
\begin{equation} \label{tp}
f(x) = \pm \sqrt{(2K^2)^{-1}\left(3-2s^2 \pm\sqrt{9-16s^2}\right)}.
\end{equation}
Assuming $ 0 < s < 3/4 $, as in classical MRI, the right-hand side yields four real values, and when $K \gg 1$ these are bunched tightly together around 0. They are thus distributed around locations where $f=0$, i.e. around the inversion points of the magnetic field. In effect, the small-scale modes are seeking out regions where the magnetic field is sufficiently weak so that tension does not stabilise them. We Taylor expand around one of the zeros of $f$, placing it at $x=0$ without loss of generality. Thus, by setting $f\approx x$ in equation \eqref{tp}, the turning points can be read off directly.

As we expect eigenmodes to be narrowly localised around the distinct zeros of $f$ and because different zeros of $f$ are separated by large evanescent zones, we treat each zero in isolation to the others. We thus set $f\approx x$ generally in the problem. 
In addition, we introduce a new space variable
$X = K x$, and equations (\ref{eq:ODE4}) and (\ref{eq:PotentialV1}) become, respectively,
\begin{align} \label{eq:FinalODE}
    \frac{d^2\Psi}{dX^2} - \frac{V(X,s)}{\epsilon^2} \Psi =0.
\end{align}
\begin{equation} \label{eq:PotentialV2}
V(X,s) = \frac{s^4+ (1+2X^2)s^2 +X^2(X^2-3)}{(s^2+X^2)^2}.
\end{equation}
Remarkably, the vertical wavenumber $K$ has dropped out entirely. If we can find growing MRI modes, then such modes can extend to arbitrarily small vertical  scales ($\sim 1/K$) and radial scales, because the wave zones have widths no greater than $1/K$.

The profile of $V$ is plotted in Fig. \ref{fig:PotentialPlot}, for a growth rate $s=0.5$. We can observe the four turning points around the zero of $f$ at $X=0$. The two positive turning points we refer to as $X_1$ and $X_2$, with $X_1 < X_2$. 
Thus, around each zero of $f$ there are five zones of differing signs of $V$. The two external zones, $|X|>X_2$, are evanescent (the eigenmode decays) and serve as barriers preventing communication between different zeros of the magnetic field.
The most internal of the five zones 
($|X|<X_1$) is also evanescent, however, as it is relatively narrow, modes may be able to tunnel through it, allowing them to inhabit both wave zones ($X_1<|X|<X_2$) concurrently.

\subsubsection{Growth rate estimates}

Equation \eqref{eq:FinalODE} can be attacked by standard Wentzel-Kramers-Brillouin-Jeffreys (WKBJ) methods, assuming $\epsilon\ll 1$. We obtain the following dispersion relation 
\begin{equation}\label{eq:DispersionRelationZNF}
\frac{1}{\epsilon} \int^{X_2}_{X_1} \sqrt{-V(t,s)} dt = \pi\left(n+\frac{1}{2}\right),
\end{equation}
where $n=0,1,2,\dots$ (see appendix \ref{appendix:asymptotics1}). Though the eigenfunctions can inhabit both wells of the potential, this tunneling introduces only an exponentially small term in the dispersion relation, which can be safely dropped. 

The fastest growing MRI mode exhibits the fewest radial oscillations and thus, in principle, the WKBJ method may be ill-suited to describe it. Nonetheless, we approximate the integral in \eqref{eq:DispersionRelationZNF} in this case. First, we set $s=3/4-\epsilon s_1$, for $s_1>0$ and small $\epsilon$. Next, $X_1$ and $X_2$ are approximated by $ \sqrt{15/16}\pm \sqrt{8s_1\epsilon/5} $, and we thus observe that the integration range is centred on $X=\sqrt{15/16}$ and is very narrow. Using this fact, we can approximate the potential as quadratic:
\begin{align*}
V= \frac{(X^2-X_1^2)(X^2-X_2^2)}{(s^2+X^2)^2}\approx \frac{5}{3}(X-X_1)(X-X_2),
\end{align*}
to leading order in $\epsilon$, which allows us to perform the integral:

\begin{equation}\label{eq:IntegralComputation}
\frac{1}{\epsilon} \int^{X_2}_{X_1} \sqrt{-V(t,s)} dt = \pi \sqrt{\frac{16}{15}}s_1.
\end{equation}

\noindent{}Combining equations (\ref{eq:DispersionRelationZNF}) and (\ref{eq:IntegralComputation}) obtains the following estimate for the fastest growing modes:

\begin{equation} \label{eq:FastestGrowingMode}
s = \frac{3}{4} - \sqrt{\frac{15}{16}}\left(n+\frac{1}{2}\right) \epsilon.
\end{equation}
This agrees surprisingly well with the numerical solutions (see Fig.~\ref{fig:GrowthRateWavenumber}). In appendix \ref{appendix:asymptotics2}, we present an alternative asymptotic calculation better suited to the fastest growing mode, which reproduces this expression exactly.

Apart from its intrinsic physical and mathematical interest, the asymptotic growth rate in equation \eqref{eq:FastestGrowingMode} is a particularly clear statement of what Fig.~\ref{fig:GrowthRateWavenumber} illustrates: near-maximal growth occurs on all sufficiently small scales. Moreover, equation \eqref{eq:FastestGrowingMode} is independent of the detailed field morphology; all that it assumes is that there exists a magnetic null. The asymptotic analysis hence demonstrates that the result holds generally.

\subsection{Conclusion}

We have shown that in the case where the background magnetic field has a zero, the linear MRI does not have a characteristic scale below which growth rates are inaccessible. On the contrary, the growth rates are significant at infinitely small scales (large wavenumbers). If it is now granted that a local linear analysis (based on the local turbulent field) meaningfully represents the work of the MRI within that self-same turbulence, then wherever the vertical field switches sign, the turbulence is not resolvable by any numerical means.  Reconnection, usually mediated by current sheets, is a critical ingredient in MHD turbulence, and thus magnetic nulls are a prevalent and essential feature of MRI turbulence, and it then follows that the resolvability issue is inescapable.





\begin{figure}
	\includegraphics[clip,trim=0 8 0 6,width=\columnwidth]{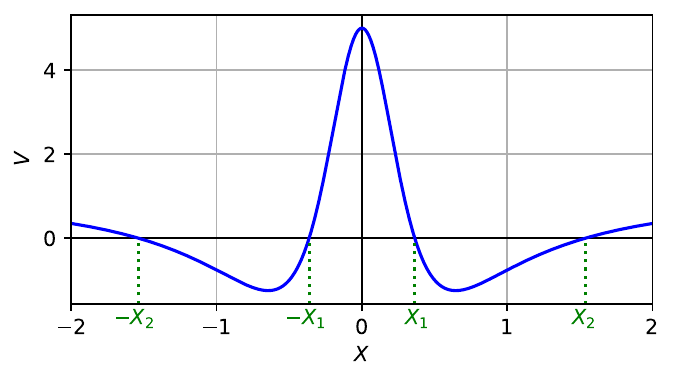}
    \caption{Potential $V$ as a function of $X = K x = k_z l_{A} x$, for a growth rate $s=0.5$. The points $\{-X_2; -X_1; X_1; X_2\}$ are the solutions of $V(X,s)=0$.}
    \label{fig:PotentialPlot}
\end{figure}

\section{Discussion: an alternative quality factor for global simulations}\label{sec:Discussions}

We now discuss a revised quality factor for global simulations that answers some, but not all, of the concerns raised in section \ref{sec:IssuesQualityFactor}. We also outline possible ways the diagnostic might be deployed and its various limitations. 

The underlying concept is relatively simple: (a) find the large-scale net vertical flux threading the disk at every cylindrical radius; (b) compute an associated quality factor, using this background field, and (c) assess its radial profile and track its evolution over time. More specifically, we calculate $B_\text{net}(R)= \langle B_z \rangle_{\upvarphi,z}$, where the average is taken over the full azimuthal extent of the domain and between $\pm n H$ vertically, with $n$ an integer (possibly 3 or 4), and $R$ is cylindrical radius. Then define the new quality factor as
 \begin{equation}\label{newcrit}
    \mathcal{Q}(R) = \frac{\lambda_\text{crit}(B_\text{net})}{\delta z} = \frac{\sqrt{3} B_\text{net}}{\sqrt{\pi \rho_\text{mid}} \Omega \delta z},
\end{equation}
where $\rho_\text{mid}$ is the midplane density (a function of cylindrical radius), and the critical MRI wavelength includes the sign of the background field. This definition aims to remove the small-scale and rapid fluctuations comprising the turbulent field, which complicated use of the traditional quality factor. Note that many global simulations use a non-uniform grid, especially when in spherical coordinates, and thus $\delta z$ will generally depend on $R$ as well.
 


The key difficulty is deciding on how large $\mathcal{Q}$ must be to obtain numerical convergence. A first approach might be to require that the local (vertically stratified) \emph{linear MRI} is sufficiently resolved at a given radius. 
 The exact threshold $\mathcal{Q}$ must meet in order to resolve this critical scale might best be determined by numerical experimentation, but we imagine that it need not be more than 10. Then equation \eqref{newcrit} can be reworked into the useful, if rough, criterion $(H/\delta z) \gtrsim \sqrt{\beta_0}$, where $\beta_0$ is the midplane plasma beta associated with the weak background net vertical flux (assumed to be a function of $R$ and $t$). 
 
 Certainly, if this criterion is satisfied, the full injection range of the turbulence will be resolved, even in the nonlinear phase (see discussion in section \ref{sec:Non-linearSaturation}). It does, however, levy a punishing demand on global simulations that few can currently meet. Moreover,
 as demonstrated by \citet{Lesur2011} and \citet{Walker2016}, the scale of maximum energy injection is larger than the scale of fastest linear MRI growth ($>\lambda_\text{crit}(B_\text{net})$). Thus any criteria based on resolving $\lambda_\text{crit}(B_\text{net})$ is likely to be overly stringent: if a simulation satisfies the criterion $\mathcal{Q} \gtrsim 10$, then it is probably resolved; but if a simulation fails the criterion, it is not necessarily unresolved.

This motivates a second approach to finding a suitable  $\mathcal{Q}$ threshold: demanding that a fixed fraction of the energy injection is numerically resolved. \cite{Walker2016} show that $\sim$80\% of the energy is injected on scales longer than $\lambda_\text{crit}(B_\text{net})$ (see their fig.~3). 
If we set our threshold at 80\%, then the criterion is again $\mathcal{Q} \gtrsim 10$. But if we set the threshold at 50\%, the less demanding criterion $\mathcal{Q} \gtrsim 2$ is perhaps sufficient, which translates to $(H/\delta z) \gtrsim 0.4 \sqrt{\beta_0}$. It must be stressed that the injection-scale estimates from \cite{Walker2016} issue from a single high-resolution simulation of fixed $\beta$; it is likely that the ratio between the nonlinear injection scale $l_{\text{nl}}$ and the background Alfv\'en length $l_{A}$ depends itself on $\beta$ (and possibly on the Reynolds numbers), though probably weakly.
Indeed, while \cite{Walker2016} show that $l_{\text{nl}}\gtrsim \lambda_\text{crit} (B_{\text{net}}) = (2 \pi / \sqrt{3}) l_A$ for $\beta  \simeq 10^3$, as $\beta \rightarrow 1$ we expect both scales to converge to $H$, and thus their ratio to approach unity.
Clearly, this requires further investigation, and we hope that future numerical work establishes a clear relationship between the nonlinear injection scale, $\lambda_\text{crit}(B_\text{net})$, and the plasma beta. If this relationship is obtained, then we can devise a tighter threshold on $\mathcal{Q}$ so as to decide on convergence.

Putting aside questions over the exact $\mathcal{Q}$ threshold, the large-scale field threading the disk, $B_\text{net}$ will presumably evolve slowly with time ($\gg \Omega^{-1}$) and vary radially on long lengthscales ($\gg H$), which answers some of the objections in section \ref{sec:SpaceVaryingMagneticFields}. In any case, this expectation can be checked throughout the simulation by plotting the radial profile of $\mathcal{Q}$ at different times. Moreover, the profiles will highlight which parts of the disk are resolved and which are less so. Additionally, imprinting the sign of $B_\text{net}$ on $\mathcal{Q}$ identifies where the global field changes polarity. Locations where $B_\text{net}$ passes through zero might be generically unresolvable (see section \ref{sec:LinearTheory}). On the other hand, if we discover that $B_\text{net}$ flips sign too rapidly with radius, then the local net-flux assumption breaks down and, as a consequence, so does the reliability of $\mathcal{Q}$. It is unclear to us what scales of radial variation are `too rapid' in this context, but we expect they fall below some critical wavelength $>H$. 

Finally, criteria based on $\mathcal{Q}$ are completely local: they assume that global MRI turbulence is a collection of local processes distributed in radius, communicating weakly with each other (see section \ref{sec:locglob}). This is unlikely to be the case in general, though it is perhaps an increasingly better approximation as $H/R$ becomes small. 

Despite the flaws and uncertainties outlined above, we believe that a criterion based on $\mathcal{Q}$ is better theoretically justified, more meaningful, and more reliable than any based on the currently used quality factor. Certainly it is worth trialling, and refining, in future global simulations.

\section{Conclusions}\label{sec:Conclusions}

In this paper we study the quality factor, commonly used to assess the convergence of the MRI in numerical simulations. We offer two detailed calculations. We show that this metric indicates zero-net-flux ideal simulations are `converged', even though they are clearly unconverged. 
This is because the Alfvén length scales not only with the grid cell size, but also with the size of our numerical domain and the disk scale height. We expect that any kind of mixed scaling, in local and global simulations, will produce large $Q_z$ irrespective of the actual convergence, or not, of the simulation.

We next develop a linear theory of the MRI in the presence of spatially varying background magnetic field, in particular one passing through zero. The quality factor's theoretical underpinning is a linear theory involving a net uniform vertical flux, and thus a well-defined minimum MRI lengthscale. But turbulent simulations exhibit rapid spatial magnetic variations. We show that when the field possesses a zero, the MRI exhibits large growth rates at infinitely small scales, and thus there is no secure minimum (linear) MRI lengthscale that can anchor the quality factor, at least as it is usually defined.  

Finally, we offer a revised quality factor for global simulations that makes use of any large-scale vertical field threading the disk, rather than the stronger, and more spatially and temporally variable, turbulent field. We discuss its better theoretical justification, but also outline its shortcomings. At the very least, in suggesting the revised metric, our hope is to stimulate a critical and informed debate on the resolvability of the MRI in global simulations and to prompt the development of more accurate diagnostics.

\section*{Acknowledgements}

The authors would like to thank the reviewer for a set of useful comments and suggestions. They are also very grateful to Geoffroy Lesur, Gordon Ogilvie, Loren Held and Jonatan Jacquemin for insightful conversations and feedback on the paper. TJ wishes to thank Josh Brown and Simon Delcamp for fruitful scientific discussions.

This research was funded by STFC (Science and Technology Facilities Council) through grant ST/X001113/1.

This work used the DiRAC Extreme Scaling service Tursa at the University of Edinburgh, managed by the Edinburgh Parallel Computing Centre on behalf of the STFC DiRAC HPC Facility (\href{www.dirac.ac.uk}{www.dirac.ac.uk}). The DiRAC service at Edinburgh was funded by BEIS, UKRI and STFC capital funding and STFC operations grants. DiRAC is part of the UKRI Digital Research Infrastructure.

The simulations were run on Tursa GPUs under DiRAC grant APP12550. Most of the post-processing was done on the Wilkes supercomputer of the Cambridge Service for Data Driven Discovery (CSD3) with Python, mainly using the libaries numpy \citep{numpy}, matplotlib \citep{matplotlib} and scipy \citep{scipy}. The numerical resolution of section \ref{sec:PositiveGrowthRatesInfinitelySmallScales} was done with MATLAB.

\section*{Data Availability}
 
Simulation data will be made available upon reasonable request to the corresponding author.


\newpage

\bibliographystyle{mnras}
\bibliography{example} 




\newpage

\appendix

\section{Quality factor histograms}\label{appendix:Histograms}

\begin{figure}
	\includegraphics[clip,trim=0 8 0 5,width=\columnwidth]{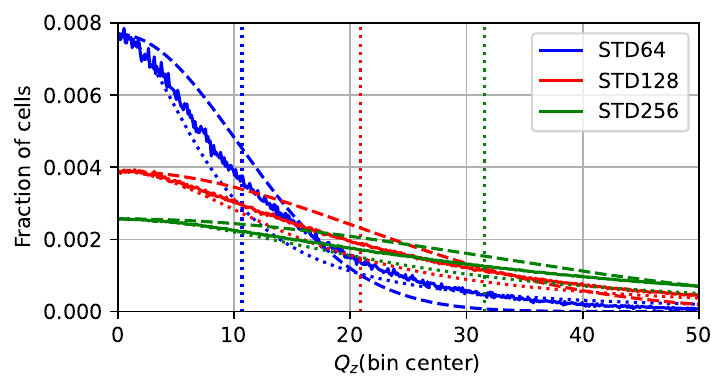}
    \caption{Histograms of $Q_z$ for simulations STD64, STD128 and STD256. The vertical lines are the mean quality factors. Dashed curves are half-Gaussian distributions and dotted curves are half-Cauchy distributions. Each color represents a different simulation.}
    \label{fig:Histogram}
\end{figure}

In this appendix, we study the statistics of the turbulence in the three simulations STD64, STD128 and STD256, using histograms of the quality factor $Q_z$. We split all possible $Q_z$ in bins of width 0.1. For Fig. \ref{fig:Histogram} we go through all simulation cells and assign their local $Q_z$ into its bin.

\begin{figure}
	\includegraphics[clip,trim=0 8 0 5,width=\columnwidth]{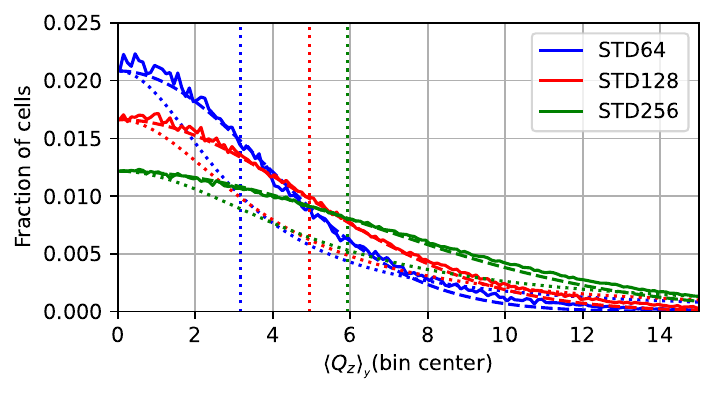}
    \caption{Histograms of $\left< Q_z \right>_y$ for simulations STD64, STD128 and STD256. The vertical lines are the mean quality factors. Dashed curves are half-Gaussian distributions and dotted curves are half-Cauchy distributions. Each color represents a different simulation.}
    \label{fig:YAveragedHistogram}
\end{figure}

For Fig. \ref{fig:YAveragedHistogram} we average $Q_z$ along the $y$-direction, then assign all $\left< Q_z \right>_y (x,z)$ to their bin. We also sum 20 different snapshots for each simulation to have reasonable statistics.

Note that as all cells have the same volume, the $y$-axis of Fig. \ref{fig:Histogram} and \ref{fig:YAveragedHistogram} show both the fraction of cells and the fraction of volume.

In both cases, the vertical lines are the means of the quality factors. The dashed curve is a half-Gaussian probability density function (PDF) and the dotted curve a half-Cauchy PDF, both normalized at $Q_z = 0$. Each color represents a different simulation (i.e. a different resolution).

The quality factor seems to lie between those two distributions, and the $y$-averaged quality factor is remarkably close to the half-Gaussian. Note however that both cases have a relatively heavy tail. The statistics of saturated ideal MRI turbulence seems to be an interesting problem. And we know that the statistics of turbulence has been studied numerically, especially to improve the understanding of the interstellar medium (see for instance \cite{Federrath2021} and references therein). But we chose not to push the analysis further, as this is out of the frame of this work.

\section{Evolution of $Q_\text{\MakeLowercase{\emph{x}}}$ and $Q_\text{\MakeLowercase{\emph{y}}}$}\label{appendix:QxQy}

\begin{figure}
	\includegraphics[clip,trim=0 10 0 6, width=\columnwidth]{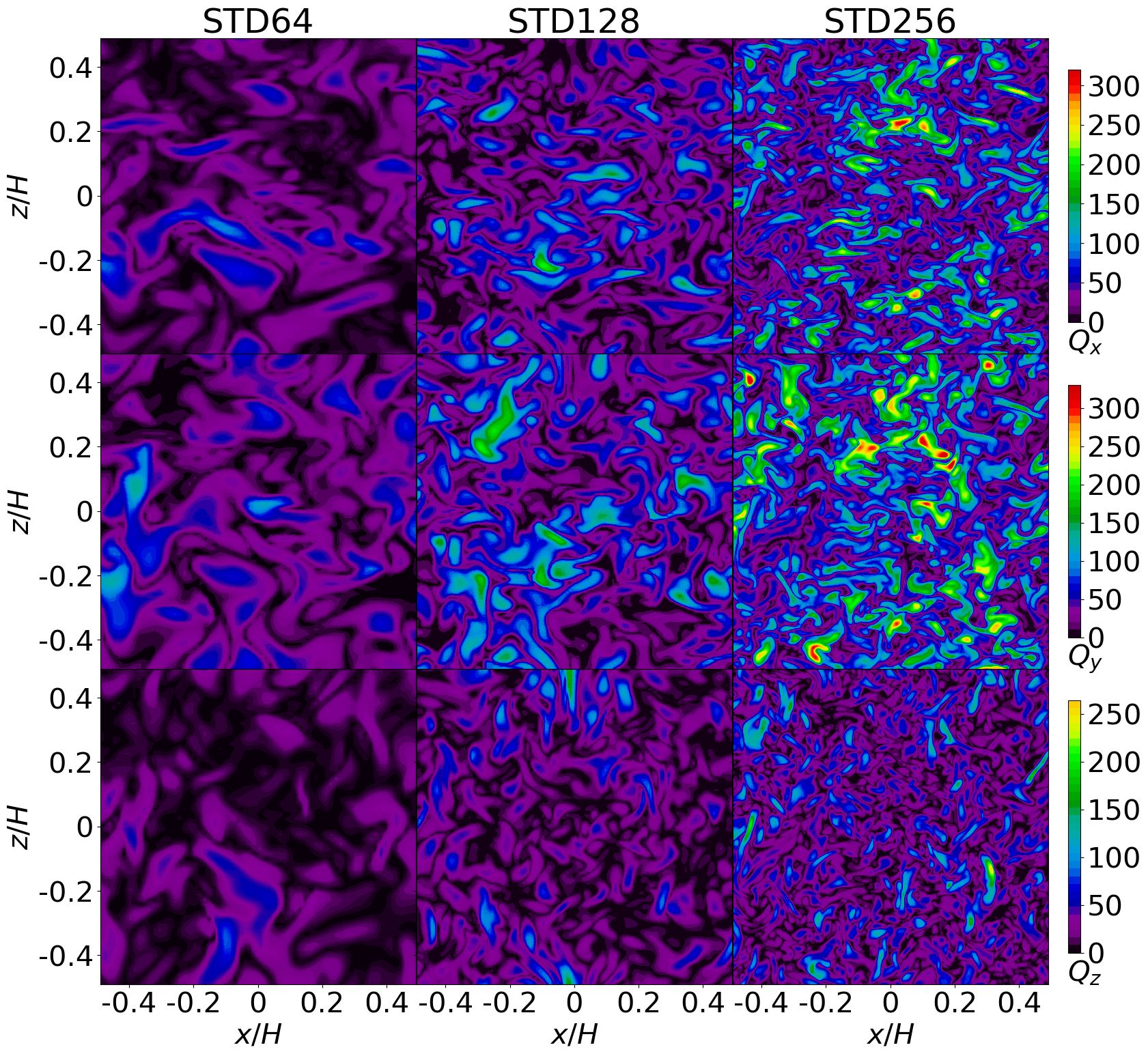}
    \caption{Distribution of $Q_x$, $Q_y$ and $Q_z$ in the $(x,z)$ plane $y=0$ for the last outputs of simulations STD64, STD128 and STD256.}
    \label{fig:CoupeAllQs}
\end{figure}

\begin{figure}
	\includegraphics[clip,trim=0 8 0 5,width=\columnwidth]{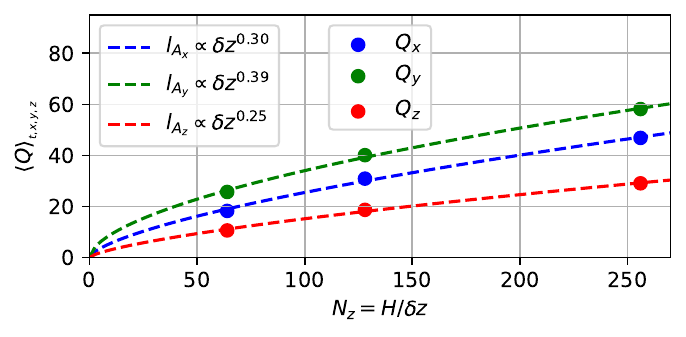}
    \caption{Points: evolution of $\left< Q_d \right>_{x,y,z}$ with resolution for $d \in \{x;y;z \}$. Lines: best fits to a monomial $l_A \propto \delta z ^p$ on those points.}
    \label{fig:ResolutionAllQs}
\end{figure}

In this appendix, we show the evolution with resolution of the quality factors along the radial direction ($Q_x$) and along the azimuthal direction ($Q_y$). They are defined as $Q_x = l_{A_x} / \delta x$ and $Q_y = l_{A_y} / \delta y$ where $l_{A_x} = \lvert V_{A_x} \rvert / \Omega$ and $l_{A_y} = \lvert V_{A_y} \rvert / \Omega$ are the radial and azimuthal Alfvén lengths, and $\delta x$ and $\delta y$ are the grid cell sizes in the radial and azimuthal directions. The azimuthal quality factor $Q_y$ is often used in conjonction with $Q_z$ (see section \ref{sec:HowItsUsedInLit}).

Fig. \ref{fig:CoupeAllQs} is analogous to Fig. \ref{fig:CoupeBy}. It shows the distributions of all three quality factors ($Q_x$, $Q_y$ and $Q_z$) in the plane $y=0$, for the last output of the three simulations STD64, STD128 and STD256. Similarly to $Q_z$, the values of $Q_x$ and $Q_y$ increase with resolution, while their characteristic scale decreases.

Fig. \ref{fig:ResolutionAllQs} is analogous to Fig. \ref{fig:ResolutionQz}. We show the evolution with resolution of the quality factors averaged over the whole box and the last 100 orbits. The lines are the best fit to a monomial $l_A \propto \delta z ^p$, where for all simulations $\delta x = \delta z \simeq \delta y$. Both $l_{A_x}$ and $l_{A_y}$ are slightly larger than $Q_z$. They both have a slightly stronger scaling with the grid cell size than $l_{A_z}$. However, that scaling is far from linear ($l_{A_x} \propto \delta z ^{0.30}$ and $l_{A_y} \propto \delta z ^{0.39}$). Thus the conclusions of section \ref{sec:LengthsOtherThanCellSize} also apply to $Q_x$ and $Q_y$, they are not good indicators of convergence either.

\section{Net-flux distributions of $B_\text{\MakeLowercase{\emph{z}}}$}\label{appendix:NetFluxBz}

\begin{figure}
    \centering
	\includegraphics[clip,trim=0 10 0 5, width=.9\columnwidth]{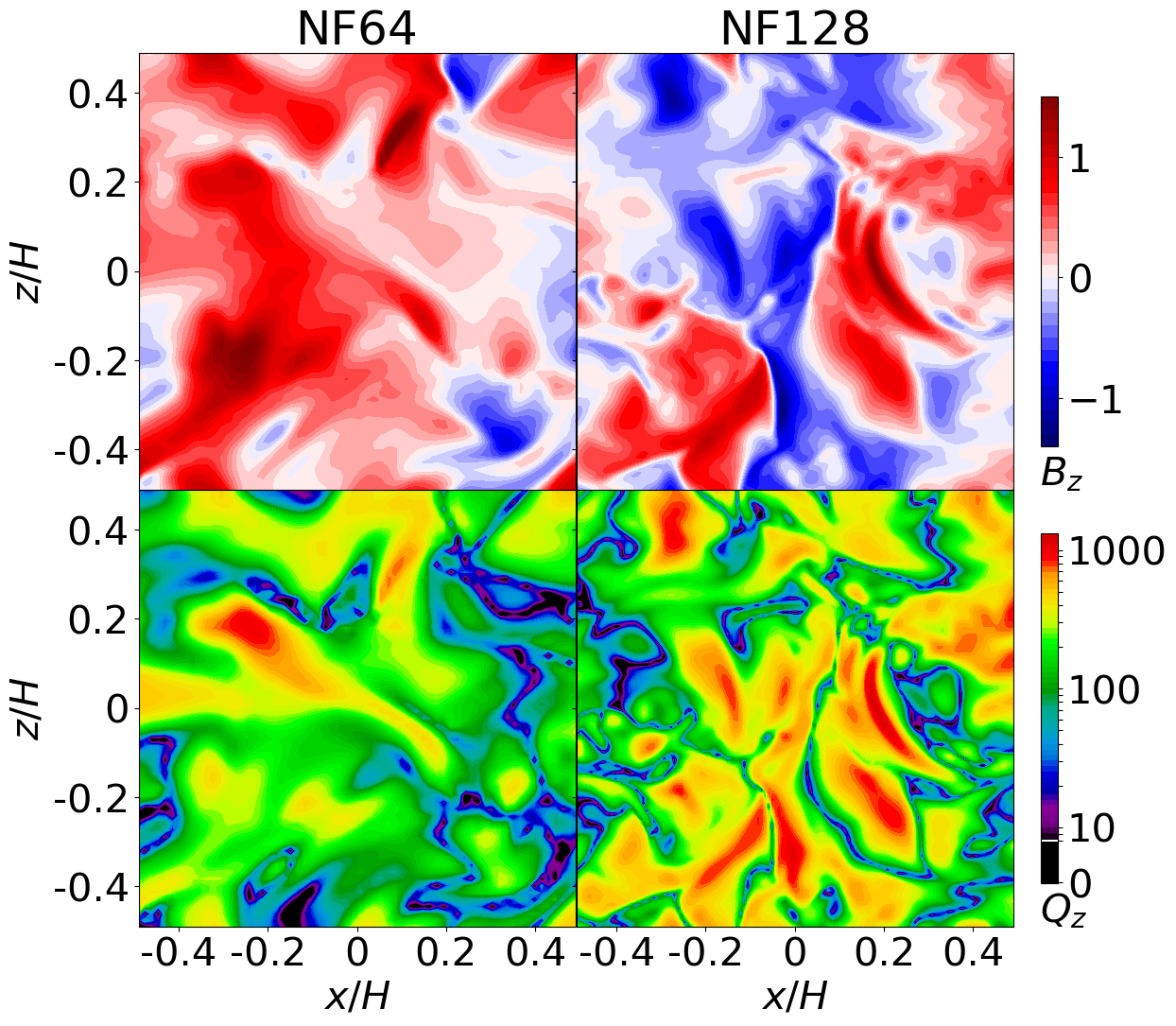}
    \caption{Distribution of $B_z$ (first row) and $Q_z$ (second row) in the $(x,z)$ plane $y=0$ for the last outputs of simulations NF64 and NF128. In the bottom row all cells with $Q_z < 8$ are shown in black.}
    \label{fig:NFCoupeBz}
\end{figure}

In this appendix, we show the distributions of $B_z$ and $Q_z$ for the net-flux simulations NF64 and NF128, represented on Fig. \ref{fig:NFCoupeBz}. As the turbulent field dwarfs the background field $B_0 / 2 \simeq 0.07$, various field reversals are observed in both simulations. On the other hand $Q_z$ is by construction always positive, and here very large.

\section{Stratified $Q_\text{\MakeLowercase{\emph{x}}}$ and $Q_\text{\MakeLowercase{\emph{y}}}$}\label{appendix:StratifiedQxQy}

\begin{figure}
	\includegraphics[clip,trim=0 8 0 6,width=\columnwidth]{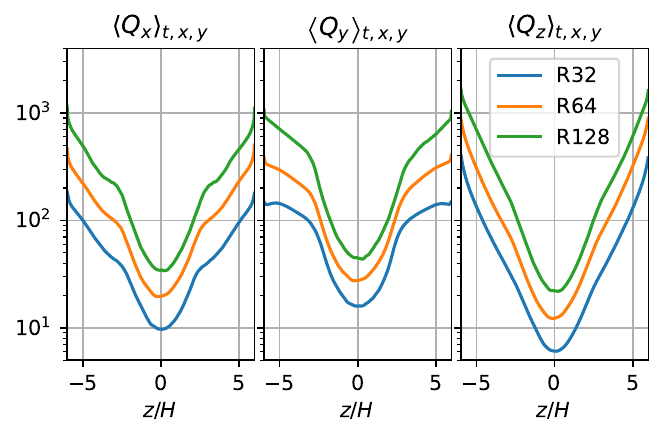}
    \caption{Vertical profiles of $ \left< Q_x \right>_{t,x,y}$ $, \left< Q_y \right>_{t,x,y}$ and $ \left< Q_z \right>_{t,x,y}$ for the stratified simulations R32, R64 and R128.}
    \label{fig:StratifiedQxyzProfiles}
\end{figure}

In this appendix, we show the vertical profiles of the quality factors for the stratified simulations (Fig. \ref{fig:StratifiedQxyzProfiles}). With increasing altitudes comes a sharp increase in all $Q$s. Yet for R128, even at the midplane $Q$s are large enough for the simulation to be deemed converged ($Q_z > 15$, $Q_y > 20$). As R128 is unconverged, this shows that no $Q$ is a good indicator of convergence for ideal local zero-net-flux simulations.

As expected, at the highest altitudes we see a dominance of $Q_z$ compared to $Q_x$ and $Q_y$, i.e. a dominance of $B_z$ compared to $B_x$ and $B_y$. Note that \cite{Simon2009} rightfully predicted that this situation could lead to the launching of \cite{blandford1982}-type winds (see \cite{Held2024} and references therein).

\section{Asymptotic approximations}\label{appendix:asymptotics}

\subsection{WKBJ method}
\label{appendix:asymptotics1}

In this appendix we present schematically the way to solve equation (\ref{eq:FinalODE}) for $\Psi$ through a WKBJ approximation. Because for each growth rate $s$ the potential is an even function of $X$, we can limit the search to $X > 0$, and look for odd and even solutions. The boundary conditions on $\Psi_{\text{odd}}(X)$ and $\Psi_{\text{even}}(X)$ are then:

\begin{itemize}
    \item Odd: $\Psi_{\text{odd}} (0) = 0$ and $\underset{X\to+\infty}{\lim} \Psi_{\text{odd}} (X)=0$.
    \item Even: $\left( d \Psi_{\text{even}}/dX \right) (0) = 0$ and $\underset{X\to+\infty}{\lim} \Psi_{\text{even}} (X)=0$.
\end{itemize}

The potential $V$ is such that $V(X) >0$ in $[0;X_1[ \cup ] X_2; + \infty [$ and $V(X) < 0$ in $]X_1;X_2[$ (see Fig. \ref{fig:PotentialPlot}). Following the usual method (see for instance chapter 10 of \cite{Bender1978}), we construct solutions of $\Psi_{\text{even}}(X)$ and $\Psi_{\text{odd}}(X)$, using sums of exponentially growing or decaying functions in $[0;X_1[$ and $]X_2;+ \infty[$, sums of Airy and Bairy functions around $X_1$ and $X_2$, and a sum of oscillatory functions in $]X_1;X_2[$. We then determine the integration constants by connecting those regions and using the boundary conditions.

This gives the following dispersion relation:
\begin{equation}\label{eq:DispersionRelationThetaPhi}
    \Theta (s) \pm \arctan \left(2 \exp{(2 \Phi (s) )}\right) = m \pi, \quad \text{with } m \in \mathds{Z} ,
\end{equation}

\noindent{}where the $+$ corresponds to the odd case and the $-$ to the even case. $\Theta(s)$ and $\Phi(s)$ are the following integrals:
\begin{equation}\label{eq:Integrals}
    \Theta (s) = \frac{1}{\epsilon} \int^{X_2}_{X_1} \sqrt{-V(t,s)} dt, \qquad \Phi (s) = \frac{1}{\epsilon} \int^{X_1}_{0} \sqrt{V(t,s)} dt .
\end{equation}

For general $s$, the integral defining $\Phi$ is independent of $\epsilon$ to leading order, and thus $\Phi\gg 1$. As a result, the second term in equation \eqref{eq:DispersionRelationThetaPhi} we take to be $\pm \pi/2$, with any associated error $\sim \exp{(-1/\epsilon)}$. Finally, noting $\Theta>0$, the dispersion relation becomes $\Theta (s) = n \pi + \pi /2$, for $n=0,1,2,\dots$, which is akin to the usual Bohr-Sommerfeld quantisation relation for a single potential well, and equivalent to equation (\ref{eq:DispersionRelationZNF}).


\subsection{Reduction to a Weber equation}
\label{appendix:asymptotics2}

Starting from equations \eqref{eq:FinalODE} and \eqref{eq:PotentialV2}, we set $s=3/4-\epsilon s_1$, as before. Assuming $\epsilon$ to be small, and cognizant of the form of $X_1$ and $X_2$ in this limit, we change the independent variable from $x$ to 
$\xi=(x-\sqrt{15/16})\epsilon^{-1/2}$. This focuses us on the outer wave zone, without loss of generality. Subsequently, we expand equation \eqref{eq:FinalODE} in $\epsilon$, and obtain the Weber equation
\begin{equation}
\frac{d^2\Psi}{d\xi^2} + \frac{1}{3}\left(8s_1-5\xi^2 \right)\Psi = 0.
\end{equation}
As indicated earlier, when approximating the WKBJ integral, conveniently the potential is a quadratic in this limit.
Next we insist our solution $\Psi$ decays for large positive $\xi$, and thus it must take the form
$\Psi \propto D_\nu ( \mu \xi ),$
where $\nu= -1/2 +\sqrt{16/15}s_1$, $\mu=(20/3)^{1/4}$, and $D_\nu$ is the parabolic cylinder function \citep{AbeSteg}. 

To determine $s_1$, strictly we should match this solution, as $\xi\to-\infty$, to an odd or even solution of equation \eqref{eq:FinalODE} near $X=0$. However, it suffices to demand $\Psi$ decays for large negative $\xi$, with any associated error exponentially small. This boundary condition is satisfied if simply $\nu=n$, where $n=0,1,2,\dots$, and we then reproduce equation \eqref{eq:FastestGrowingMode}.

\bsp	
\label{lastpage}
\end{document}